\documentstyle[11pt,oldlfont]{article}
\def\theequation{1.\arabic{equation}}
\normalbaselines
\begin{document}
\bibliographystyle{unsrt}

\title{\LARGE  SQUEEZED AND CORRELATED STATES\\
       OF QUANTUM FIELDS AND  MULTIPLICITY\\
       PARTICLE DISTRIBUTIONS}
\author{\Large V.V. Dodonov, I.M. Dremin, O.V. Man'ko,\\
            \Large V.I. Man'ko, and P.G. Polynkin\\
                   Lebedev Physical Institute,\\
                   Leninsky Prospekt, 53, 117924, Moscow, Russia}
\date{}
\maketitle
\begin{abstract}
The primary aim of the present paper is to attract the attention of particle
physicists to new developments in studying squeezed and correlated states
of the electromagnetic field as well as of those working on the latest topic
to new findings about multiplicity distributions in quantum chromodynamics.
The new types of nonclassical states used in quantum optics as squeezed
states, correlated states, even and odd coherent states (Schr\"odinger cat
states) for one--mode and multimode interaction are reviewed.
Their distribution functions are analyzed according to the method used first
for multiplicity distributions in high energy particle interactions.
The phenomenon of oscillations of particle distribution functions of the
squeezed fields is described and confronted to the phenomenon of oscillations
of cumulant moments of some distributions for squeezed and correlated field
states. Possible extension of the method to fields different from the
electromagnetic field (gluons, pions, etc.) is speculated.
\end{abstract}
\vspace{1.5cm}



\newpage

\voffset=-1.5cm
\setcounter{equation}{0}
\begin{center}
{\Large 1. INTRODUCTION}
\end{center}
The nature of any source of radiation (of photons, gluons or other
entities) can be studied by analyzing multiplicity distributions,
energy spectra, various correlation properties, etc. A particular
example is provided by the coherent states of fields which give rise to
Poisson distribution. However, in most cases one has to deal with various
distributions revealing different dynamics. We try to describe some of them
appearing in electrodynamics and in chromodynamics but start first with
coherent states. As incomplete as it is, this review, we hope, will stimulate
physicists working on different topics to think over similarity of their
problems and to look for common solutions.

The coherent states for photons have been introduced in \cite{Glauber63}.
They are widely used in electrodynamics and quantum optics
\cite{SudarshanKla,KlaSkag}.
The coherent states of some nonstationary quantum systems were
constructed in \cite{MalTri69}-\cite{MalManTri73},
and this construction was based on finding some new
time--dependent integrals of motion. The integral of motion which is
quadratic in position and momentum was found for classical oscillator
with time--dependent frequency a long time ago by Ermakov \cite{Er1880}.
His result was rediscovered by Lewis \cite{Lewis}.
The two time--dependent integrals of motion which are linear forms in
position and momentum for the classical and quantum oscillator with
time--dependent frequency, as well as corresponding coherent states
were found in \cite{MalMan70};
for a charge moving in varying in time uniform magnetic field this was done
in \cite{MalTri69} (in the absense of the electric field),
\cite{MalJETP,MalTriJETP} (nonstationary magnetic field with the
``circular'' gauge of the vector potential plus uniform nonstationary
electric field), and \cite{MalDod72} (for the Landau gauge of the
time-dependent vector potential plus nonstationary elecrtic field).
 For the multimode nonstationary oscillatory
systems such new integrals of motion, both of Ermakov's type (quadratic in
positions and momenta) and linear in position and momenta, generalizing the
results of \cite{MalMan70} were constructed in \cite{MalManTri73}.
The approach of constructing the coherent
states of the parametric systems based on finding the time--dependent
integrals of motion was reviewed in \cite{MalMan79} and \cite{DodMan89}.
Recently the discussed time--dependent invariants were obtained using
Noether's theorem in \cite{Cas}-\cite{Castanes}.

The coherent states \cite{Glauber63} are considered  (and called) as classical
states of the field. This is related to their property of equal noise in
both quadrature components (dimensionless position and momentum)
corresponding to the noise in the vacuum state of the electromagnetic field,
and to the minimization of the Heisenberg uncertainty relation \cite{Heis27}.
These two properties yield the Poisson particle distribution function in the
field coherent state, which is the characteristic feature of the
classical state.

On the other hand, the so--called {\it squeezed states} of the photons were
considered in quantum optics for the one--mode field
\cite{Stol}-\cite{Walls}. The main property of these states is that
the quantum noise in one of the field quadrature components is less than for
the vacuum state. The important feature of the nonclassical states is the
possibility to get statistical dependence of the field quadrature components
if their correlation coefficient  is not equal to zero. For the classical
coherent state and for the particle number state there is no such
correlation. The correlated states possessing the quadrature statistical
dependence were introduced in \cite{Kur80} by usage of the procedure of
minimization of the Schr\"odinger--Robertson uncertainty relation
\cite{Sch30,Rob30}. The nonclassical states of another type, namely,
even and odd
coherent states, were introduced in \cite{DM74} and called as
the Schr\"odinger cat states \cite{Yur86}. The even and odd coherent states are
very simple even and odd superpositions of the usual coherent states. The
particle distribution in these states differs essentially from the Poissonian
statistics of the classical states. Its most striking feature is the
oscillations of the particle distribution functions which is characteristic
property of nonclassical light \cite{WeinVourdas}-\cite{Rosa}.
They are especially strong for even and odd
coherent states \cite{DM74,Nikonov93}, moreover, they are very sensitive to the
correlation of the quadrature components \cite{Rosa}. These phenomena are
typical not only for the one--mode photons production, but also for the
multimode case \cite{Ola1}-\cite{Ansari94}.
The multimode generalizations of the Schr\"odinger cat states
were studied in \cite{Nikonov93,Ansari94}.

Let us turn now to high energy particle interactions. Several years ago the
experimentalists of UA5 Collaboration in CERN noticed \cite{UA5} a shoulder in
the multiplicity distribution of particles produced in $p\bar p$ collisions at
energies ranging from 200 to 900 GeV in the center of mass system.
It looked like a small wiggle over a smooth curve and was immediately ascribed
by theorists to processes with larger number of
Pomerons exchanged in the traditional schemes. More recently, several
collaborations studying $e^{+}e^{-}$ collisions at 91 GeV in CERN  reported
(see, e.g., \cite{DELPHI,OPAL}) that they failed to fit the multiplicity
distributions of produced particles by smooth curves (the Poisson and Negative
Binomial distributions were among them). Moreover, subtracting such smooth
curves from the experimental ones they found steady oscillatory behaviour of
the difference. It was ascribed to the processes with different number of jets.

In such circumstances one is tempted to speculate about the alternative
explanation when considering
possible similarity of these findings to typical features of squeezed and
correlated states. Indeed, we know that the usual coherent states appeared
important for the theory of particle production \cite{Botke}. Moreover,
the squeezed states, being introduced initially
for solving the problems of quantum optics, now begin to penetrate to
different other branches of physics, from solid state physics
\cite{Art}-\cite{Kovar} to cosmology and gravitation \cite{Sidor}-\cite{Albr}.
Thus why they could not arise in particle physics?
In slightly different context the ideas about squeezed
states in particle physics were promoted in
\cite{VourdasWeiner}-\cite{Belorus}. No attempts to use analogy with
other types of nonclassical
states like Schr\"odinger cat states or correlated states are known.

In the above approach to experimental data, however, the form of oscillations
depends on the background subtracted. The new sensitive method
was proposed in \cite{Dremin,DreminPhysLett} (for the review
see \cite{DreminUFN}).
It appeared as a byproduct of the solution of the equations for generating
functions of multiplicity distributions in quantum chromodynamics (QCD).
It appeals to the moments of the multiplicity distribution. According to QCD,
the so--called cumulant moments (or just cumulants), described in more
details below, should reveal the oscillations as functions of their ranks,
while they are identically equal to zero for the Poisson distribution
and are steadily decreasing
functions for Negative Binomial distribution so widely used in
phenomenological analysis. Experimental data show the oscillatory behaviour
of cumulants (see \cite{Gian}) of multiplicity distributions in high energy
inelastic processes initiated by various particles and nuclei,
even though some care should be taken due to
the high multiplicity cut-off of the data. When applied to the squeezed states,
the method demonstrates \cite{Pavel} the oscillations of cumulants in slightly
squeezed states and, therefore, can be useful for their detection.

As we have already pointed out above, the aim of this article is to review
the properties of the nonclassical
states (squeezed, correlated, even and odd coherent ones) both for one- and
multimode cases in the context of possible applications of them to high
energy physics, as well as to acquaint with new methods of analysis of
multiplicity distributions. However, we will not speculate here about
any application in more details,
leaving the topic for future publications.

The article is organized as follows. In sections 2 and 3 we describe the
quantities used in analysis of multiplicity distributions for
one--dimensional and multidimensional cases, respectively. In sections 4 and
5 we introduce the quadrature components of boson fields, describe their
properties and discuss their connection with experimentally measured
quantities for electromagnetic field. In section 6 the formalism of Wigner
function and Weyl transformation is briefly discussed.
In section 7 we give the explicit formulas describing the polymode mixed
boson field (photons, phonons, pions, gluons, and etc.).
Section 8 is dedicated to studying the one--mode squeezed and correlated
light.  We discuss the beha the photon distribution function and of
its moments and demonstrate their strongly oscillating character for the
slightly squeezed states of a field. Even and odd coherent states of
one--mode boson field are considered in section 9.
Concluding remarks are given in section 10.


\setcounter{equation}{0}
\def\theequation{2.\arabic{equation}}

\begin{center}
{\Large 2. DISTRIBUTION FUNCTION AND ITS MOMENTS.\\ ONE--DIMENSIONAL CASE}
\end{center}

In this part of the review we will briefly summarize definitions and
notations for the values that are used to characterize various processes
of inelastic scattering according to number of particles produced in these
processes. We will also point out the relations connecting these values to
each other and give the examples related to some distributions typical in the
probability theory.

Any process of inelastic scattering (that is the scattering with new
particles produced) can be characterized by the function $P_{n}$, the
multiplicity distribution function. The value of $P_{n}$ denotes the
probability to observe $n$ particles produced in the collision. It
is clear that $P_{n}$ must be normalized to unity:
\begin{eqnarray}
  \sum_{n=0}^{\infty} P_{n}=1.
\end{eqnarray}
Sometimes the multiplicity distribution of particles produced can be
conveniently described by its moments. It means that the series of numbers
$P_{n}$ is replaced by another series of numbers according to a certain rule.
All these moments can be obtained by the differentiation of the so--called
generating function $G(z)$ defined by the formula:
\begin{eqnarray}
  G(z)=\sum_{n=0}^{\infty} P_{n} z^{n}.
\end{eqnarray}
Thus, instead of the discrete set of numbers $P_{n}$ we can study the
analytical function $G(z)$.

We will use the factorial moments and cumulants defined by the following
relations:
\begin{eqnarray}
    &&F_{q}=\frac {\sum_{n=0}^{\infty }P_{n}(n-1)\cdots (n-q+1)}
    {\left(\sum_{n=0}^{\infty }P_{n}n\right)^{q}}
    =\left.\frac{1}{\langle n \rangle^{q}}\frac{d^{q}G(z)}{dz^{q}}
    \right|_{z=1},
\end{eqnarray}
\begin{eqnarray}
  \left.K_{q}=\frac{1}{\langle n \rangle^{q}}\frac{d^{q}\mbox{ln}G(z)}
  {dz^{q}} \right|_{z=1},
\end{eqnarray}
where
\begin{eqnarray}
  \langle n \rangle =\sum_{n=0}^{\infty }P_{n}n
\end{eqnarray}
is the average multiplicity.

Reciprocal formulas expressing the generating function in terms of cumulants
and factorial moments can also be obtained:
\begin{eqnarray}
  \lefteqn{}
    &&G(z)=\sum_{q=0}^{\infty} \frac{z^{q}}{q!} \langle n \rangle^{q} F_{q}
    \nonumber\\
    &&(F_{0}=F_{1}=1),
\end{eqnarray}
\begin{eqnarray}
  \lefteqn{}
    &&\mbox{ln}G(z)=\sum_{q=1}^{\infty} \frac{z^{q}}{q!} \langle n
    \rangle^{q} K_{q} \nonumber\\
    &&(K_{1}=1).
\end{eqnarray}
The probability distribution function itself is related to the generating
function in the following way:
\begin{eqnarray}
  \left. P_{n}=\frac{1}{n!}\frac{d^{n}G(z)}{dz^{n}} \right|_{z=0}.
\end{eqnarray}
Factorial moments and cumulants are connected to each other by the following
recursion relation:
\begin{eqnarray}
  F_{q}=\sum_{m=0}^{q-1}C_{q-1}^{m} K_{q-m} F_{m},
\end{eqnarray}
where
\begin{eqnarray}
  C_{q-1}^{m} = \frac{(q-1)!}{m! (q-m-1)!}
  \nonumber
\end{eqnarray}
are binomial coefficients.
This formula can be easily obtained if we remind that $F_{q}$ and $K_{q}$ are
defined as the $q$--th order derivatives of the generating function and its
logarithm, respectively, and use the well--known formula for differentiation
of the product of two functions:
\begin{eqnarray}
  \lefteqn{}
    &&\sum_{m=0}^{q-1}C_{q-1}^{m} K_{q-m} F_{m} \nonumber\\
    &&=\sum_{m=0}^{q-1}C_{q-1}^{m}\left[\frac{1}{\langle n \rangle ^{q-m}}
      \frac{d^{q-m}}{dz^{q-m}}\mbox{ln}G(z)\frac{1}{\langle n \rangle ^{m}}
      \frac{d^{m}}{dz^{m}} G(z) \right] _{z=1} \nonumber\\
    &&=\frac{1}{\langle n \rangle ^{q}}\sum_{m=0}^{q-1} C_{q-1}^{m} \left[
      \frac{d^{q-1-m}}{dz^{q-1-m}}\frac{d/dz \hspace{0.1cm} G(z)}{G(z)}
      \frac{d^{m}}{dz^{m}}G(z) \right]_{z=1} \nonumber\\
    &&=\frac{1}{\langle n \rangle ^{q}}
      \left[\frac{d^{q-1}}{dz^{q-1}}\left(\frac{d/dz
      \hspace{0.1cm} G(z)}{G(z)}G(z)\right) \right] _{z=1} \nonumber\\
    &&=\left. \frac{1}{\langle n \rangle ^{q}}
      \frac{d^{q-1}}{dz^{q-1}}d/dz \hspace{0.1cm}G(z) \right| _{z=1}=F_{q}.
\end{eqnarray}
Relation (2.9) gives the opportunity to find the factorial moments if
cumulants are known, and vice versa.

It must be pointed out that cumulants are very sensitive to small variations
of the distribution function and hence can be used to distinguish the
distributions which otherwise look quite similar.

Usually, cumulants and factorial moments for the distributions occuring in
the particle physics are very fast growing with increase of their rank.
Therefore sometimes it is more convenient to use their ratio:
\begin{eqnarray}
  H_{q}=\frac{K_{q}}{F_{q}} ,
\end{eqnarray}
which behaves more quietly with increase of $q$ remaining a
sensitive measure of the tiny details of the distributions.

In what follows we imply that the rank of the distribution function moment is
non--negative integer even though formulas (2.3) and (2.4) can be generalized
to the non--integer ranks.

Let us demonstrate two typical examples of the distribution.

$1.\hspace{0.5cm} Poisson \hspace{0.4cm} distribution.$

The Poisson distribution has the form:
\begin{eqnarray}
  P_{n}=\frac{\langle n \rangle^{n}}{n!}\mbox{exp}\left(-\langle n \rangle
  \right).
\end{eqnarray}
Generating function (2.2) can be easily calculated:
\begin{eqnarray}
  G(z)=\mbox{exp}\left(\langle n \rangle z\right).
\end{eqnarray}
According to (2.3) and (2.4) we have for the moments of this distribution:
\begin{eqnarray}
  F_{q}=1, \hspace{0.5cm} K_{q}=H_{q}=\delta_{q1}.
\end{eqnarray}

$2.\hspace{0.5cm}Negative\hspace{0.4cm}binomial\hspace{0.4cm}distribution.$

This distribution is rather successfully used for fits of main features of
experimental data in particle physics. It has the form:
\begin{eqnarray}
  P_{n}=\frac{\Gamma(n+k)}{\Gamma(n+1)\Gamma(k)}\left(\frac{\langle n \rangle}
  {k}\right)^{n}\left(1+\frac{\langle n \rangle}{k}\right)^{-(n+k)},
\end{eqnarray}
where $\Gamma$ is the gamma function, and $k$ is a fitting parameter.

At $k=1$ we have the usual Bose distribution. The Poisson distribution can be
obtained from (2.15) in the limit $k\rightarrow \infty$.

Generating function for the negative binomial distribution reads:
\begin{eqnarray}
  G(z)=\left( 1-\frac{z\langle n \rangle}{k} \right)^{-k},
\end{eqnarray}
and the moments of this distribution are:
\begin{eqnarray}
  \lefteqn{}
    &&F_{q}=\frac{\Gamma(k+q)}{\Gamma(k)k^{q}}, \nonumber\\
    &&K_{q}=\frac{\Gamma(q)}{k^{q-1}}, \nonumber\\
    &&H_{q}=\frac{\Gamma(q)\Gamma(k+1)}{\Gamma(k+q)}.
\end{eqnarray}

\newpage

\setcounter{equation}{0}
\def\theequation{3.\arabic{equation}}

\begin{center}
{\Large 3. DISTRIBUTION FUNCTION. \\
  $N$--DIMENSIONAL CASE}
\end{center}

Suppose that in a process of inelastic collision the particles of $N$
different species appear. Such a situation can be met, for example, in
considering the electromagnetic field, when these ``particles of different
species'' are photons corresponding to the different modes of a field. To
characterize the process of such a kind it is necessary to introduce the
values analogous to those introduced in the previous section generalizing
them to the multidimensional case.

To abbreviate the formulas we introduce the following notations:
\begin{eqnarray}
{\bf n}=(n_{1},n_{2},\ldots n_{N}) \hspace{0.4cm} - \nonumber
\end{eqnarray}
$N$--dimensional vector with integer non--negative components;
\begin{eqnarray}
{\bf \alpha }=(\alpha_{1},\alpha_{2},\ldots \alpha_{N}) \hspace{0.4cm} -
\nonumber
\end{eqnarray}
$N$--dimensional vector with complex components;
\begin{eqnarray}
{\bf z}=(z_{1},z_{2},\ldots z_{N}) \hspace{0.4cm} - \nonumber
\end{eqnarray}
$N$--dimensional vector with complex components;
\begin{eqnarray}
  \lefteqn{}
    &&{\bf n}!=n_{1}!n_{2}!\cdots n_{N}!; \nonumber\\
    &&{\bf \alpha^{n}}=\alpha_{1}^{n_{1}}\alpha_{2}^{n_{2}}\cdots
      \alpha_{n}^{n_{N}};
      \nonumber\\
    &&{\bf z^{n}}=z_{1}^{n_{1}}z_{2}^{n_{2}}\cdots z_{N}^{n_{N}};
      \nonumber\\
    &&\sum_{{\bf n}=0}^{\infty}=\sum_{n_{1}=0}^{\infty}
      \sum_{n_{2}=0}^{\infty}\cdots \sum_{n_{N}=0}^{\infty}.
\end{eqnarray}
Suppose that the particles of $N$ species are obtained as a result of some
process. Similar to the one--dimensional case, such a process can be
characterized by the multiplicity distribution function $P_{\bf n}$.
The value of $P_{\bf n}$ yields the probability to observe $n_{1}$
particles of the first kind, $n_{2}$ of the second one, and so on, up to $N$.
$P_{\bf n}$ is normalized:
\begin{eqnarray}
  \sum_{{\bf n}=0}^{\infty} P_{\bf n}=1.
\end{eqnarray}
The mean number of the $i$--th sort of particles can be obtained from
$P_{\bf n}$ in the following way:
\begin{eqnarray}
  \langle n_{i} \rangle =\sum_{{\bf n}=0}^{\infty} P_{\bf n}
   n_{i}.
\end{eqnarray}
The mean number of particles is obtained by summation of (3.3)
over the index $i$:
\begin{eqnarray}
  \langle n_{total} \rangle =\sum_{i=1}^{N} \sum_{\bf{n}=0}^{\infty}
      P_{\bf n} n_{i}=
      \sum_{{\bf n}=0}^{\infty}\left( n_{1}+n_{2}+\cdots +n_{N}\right)
      P_{\bf n}.
\end{eqnarray}
Just like for the one--dimensional case, instead of set of numbers
$P_{\bf n}$, we can analyze the analytical function $G({\bf z})$
analogous to generating function $G(z)$ introduced in (2.2):
\begin{eqnarray}
  G({\bf z})=\sum_{{\bf n}=0}^{\infty} P_{\bf n}
  {\bf z}^{\bf n}.
\end{eqnarray}
{}From the definition (3.5) it is clear that
\begin{eqnarray}
  \left. P_{\bf n}=\frac{1}{{\bf n}!}
  \frac{\partial^{n_{1}}}{\partial z_{1}^{n_{1}}}
  \cdots \frac{\partial^{n_{N}}}{\partial z_{N}^{n_{N}}}
  G\left({\bf z}\right)\right|_{{\bf z}=0}.
\end{eqnarray}
In principle, for more detailed analysis of distributions, in the
multidimensional case the cumulant and factorial moments can be defined in
a way analogous to the method used in the previous section
\begin{eqnarray}
  \lefteqn{}
    &&\left. K_{\bf n}=\frac{1}{\langle n_{1} \rangle \cdots
    \langle n_{2}\rangle} \frac{\partial^{n_{1}}}{\partial z_{1}^{n_{1}}}
    \cdots \frac{\partial^{n_{N}}}{\partial z_{N}^{n_{N}}}
    \mbox{ln}G({\bf z})\right|_{{\bf z}=1}, \nonumber\\
    &&\left. F_{\bf n}=\frac{1}{\langle n_{1} \rangle \cdots
    \langle n_{2}\rangle} \frac{\partial^{n_{1}}}{\partial z_{1}^{n_{1}}}
    \cdots \frac{\partial^{n_{N}}}{\partial z_{N}^{n_{N}}}
    G({\bf z})\right|_{{\bf z}=1}.
\end{eqnarray}
However, these values, unlike multiplicity distribution function
$P_{\bf n}$ and generating function $G({\bf z})$, are not
so widely used.
\setcounter{equation}{0}
\newpage

\def\theequation{4.\arabic{equation}}

\begin{center}
{\Large 4. QUADRATURE COMPONENTS.\\
   ONE--MODE CASE}
\end{center}

Many fields, appearing in physics for description of different interactions,
can be quantized according to the boson scheme. As examples we mention here
the electromagnetic field, the field of oscillations in a crystal lattice,
the field of strong interactions, and so on. The technique of such
quantization is well--known \cite{Ahiezer}. In fact, it appeals to reducing
a field to the set of noninteracting harmonic oscillators. In this section
we will give a definition of so--called quadrature components -- the
variables important for the theory. We will also discuss their connection
with real physical quantities, measured in an experiment, for the
electromagnetic field.

As we already stated above, the quantization of a field means reducing
it to the set of harmonic oscillators. Each such an oscillator corresponds,
as one says, to one mode of a field. In this part of the review we consider
the so--called one--mode field, namely the field described in complete analogy
with a single harmonic oscillator.

According to the usual practice we start with the pair of boson creation and
annihilation operators with the commutation rule:
\begin{eqnarray}
  \left[ \hat{a}, \hat{a}^{+} \right]=1.
\end{eqnarray}
We consider here the stationary case in the Schr\"odinger representation.
Therefore $\hat{a}$ and $\hat{a}^{+}$ are constant operators (not depending
on time). By definition, we introduce the quadrature components of a field.
In the case of one--mode field there are two of them:
\begin{eqnarray}
  \hat{x}=\frac{\hat{a}+\hat{a}^{+}}{\sqrt{2}}, \hspace{0.5cm}
  \hat{p}=\frac{\hat{a}-\hat{a}^{+}}{i \sqrt{2}}.
\end{eqnarray}
The following commutation rule for operators $\hat{x}$ and $\hat{p}$ is the
immediate consequence of relation (4.1) and definition (4.2):
\begin{eqnarray}
  \left[\hat{x}, \hat{p} \right]=i.
\end{eqnarray}
Situation here is entirely analogous to that of a usual harmonic oscillator.
The only difference is the meaning of operators $\hat{x}$ and $\hat{p}$.
These operators in a general case are not necessarily the usual coordinate and
momentum. The real physical sense of quadrature components is clarified when
both the nature and the form of the field under consideration is specified.
For example, for the electromagnetic field, the role of quadrature components
is played by the vector potential and the electric field intensity.

What is the meaning of the quadrature component's name? To answer this
question consider classical field oscillations, the physical nature of which
is of no importance for us here. It is well--known that every value,
harmonically oscillating with time, can be expressed in the form:
\begin{eqnarray}
  A(t)=\frac{1}{\sqrt{2}}\left(a e^{-i\omega t}+a^{*} e^{i\omega t}\right),
\end{eqnarray}
where $a=\frac{1}{\sqrt{2}}(x+ip),$  $a^{*}$ is complex conjugated to $a$,
$x$ and $p$ are constant real numbers, $\omega$ is the frequency of
oscillations. Then from (4.4) it immediately follows that
\begin{eqnarray}
  A=x\hspace{0.15cm}\mbox{cos}(\omega t)+p\hspace{0.15cm}\mbox{sin}(\omega t),
\end{eqnarray}
where
\begin{eqnarray}
  x=\frac{1}{\sqrt{2}}\left(a+a^{*}\right), \hspace{0.5cm}
  p=\frac{1}{i\sqrt{2}}\left(a-a^{*}\right). \nonumber
\end{eqnarray}
$x$ and $p$ are called the ``quadrature components'' by virtue of
relation (4.5). The point is that the phase difference between sine and
cosine is $\pi/2$, and the value of $a$ can be represented as a
hypotenuse of right triangle with legs equal to the values $x$ and $p$.
Under canonical quantization of such an oscillation the values $a$,
$a^{*}$, $x$, $p$, $A$ are replaced by operators, but the name
``quadrature components'' for operators corresponding to $x$ and $p$
remains.

Now turn our attention to the problem of describing the field state. Every
quantum system (in particular, the mode of a field) can be described in terms
of state vector $|\psi \rangle$ in the case of pure state, when the system is
isolated, and in terms of density operator $\hat{\rho}$, if the state is
mixed and our system is a part (subsystem) of some larger system. The pure
state can be considered as a particular case of the mixed one with
$\hat{\rho}=|\psi \rangle \langle \psi|$.

Thus, to describe the state of the mode of a field, we can explicitly specify
the state vector $|\psi \rangle$ in some representation for the pure state,
or all matrix elements of density operator $\hat{\rho}$ in the case of mixed
state.

However, the different approach is also possible. Every state of a field can be
characterized by the mean values (moments) of quadrature components in this
state. The first order moments are the quantum mean values of operators
$\hat{x}$ and $\hat{p}$:
\begin{eqnarray}
  \lefteqn{}
    &&\langle\hat{x} \rangle =\mbox{Tr}\left(\hat{\rho}\hat{x} \right),
      \nonumber\\
    &&\langle\hat{p} \rangle =\mbox{Tr}\left(\hat{\rho}\hat{p} \right).
\end{eqnarray}
The second order moments (dispersions and covariances) are defined by
relations:
\begin{eqnarray}
  \lefteqn{}
    &&\sigma_{xx}=\langle {\hat{x}}^{2} \rangle -\langle \hat{x} \rangle
      ^{2} =
      \mbox{Tr}\left( \hat{x} ^{2}\hat{\rho}\right)-\langle
      \hat{x}\rangle^{2}, \nonumber\\
    &&\sigma_{pp}=\langle
      \hat{p}^{2}\rangle -\langle\hat{p}\rangle^{2}=
      \mbox{Tr}\left(\hat{p}^{2}\hat{\rho}\right)-\langle \hat{p}\rangle^{2},
      \nonumber\\
    &&\sigma_{xp}=\sigma_{px}=\frac{1}{2}
      \mbox{Tr}~\left[\left(\hat{x}\hat{p}+\hat{p}\hat{x}\right)\hat{\rho}
      \right]-\langle \hat{x} \rangle \langle \hat{p} \rangle .
\end{eqnarray}
{}From the commutation rule (4.3) it follows that the uncertainty relation must
be valid for any state:
\begin{eqnarray}
  \sigma_{xx} \sigma_{pp} \geq \frac{1}{4}.
\end{eqnarray}
As usual, to perform calculations, we have to fix the basis in a space of
states. The natural basis is that of the Fock states $|n \rangle$, obtained
from the vacuum state $|0 \rangle$ defined by formula
\begin{eqnarray}
  \hat{a} |0 \rangle =0,
\end{eqnarray}
with $n$--multiple action of creation operator $\hat{a}^{+}$:
\begin{eqnarray}
  |n \rangle =\frac{\hat{a}^{+n}}{\sqrt{n!}} |0 \rangle.
\end{eqnarray}
The vector $|n \rangle$ corresponds to the state of a field with $n$ quanta
of oscillations. It is the eigenvector with respect to the operator of
number of particles $\hat{N}=\hat{a}^{+}\hat{a}$:
\begin{eqnarray}
  \hat{N} |n \rangle =n |n \rangle.
\end{eqnarray}
The Fock states form the orthonormalized system:
\begin{eqnarray}
  \langle n|m \rangle =\delta_{nm}.
\end{eqnarray}
Another convenient basis is the basis of coherent states. By definition,
a state $|\alpha \rangle$ is the coherent one, if it is the eigenstate of
annihilation operator $\hat{a}$:
\begin{eqnarray}
  \hat{a}|\alpha \rangle =\alpha |\alpha \rangle.
\end{eqnarray}
Coherent states are parametrized by the continuous complex parameter
$\alpha$. They can be expressed through the Fock states:
\begin{eqnarray}
  |\alpha \rangle =\mbox{exp}\left(-\frac{|\alpha|^{2}}{2}\right)
  \sum_{n=0}^{\infty} \frac{\alpha^{n}}{\sqrt{n!}} |n \rangle.
\end{eqnarray}
The uncertainty relation (4.8) becomes minimized in coherent states:
\begin{eqnarray}
  \sigma_{xx} \sigma_{pp} =\frac{1}{4},
\end{eqnarray}
which is the most important property of these states. Coherent states
are studied in detail and widely described in literature
(see \cite{SudarshanKla,KlaSkag,MalMan79}).

The distribution function $P_{n}$, discussed in section 2, is defined
by formula:
\begin{eqnarray}
  P_{n}=\mbox{Tr}\left(\hat{\rho}|n \rangle \langle n| \right).
\end{eqnarray}
Rewrite this formula using the Fock basis:
\begin{eqnarray}
  P_{n}=\sum_{m=0}^{\infty} \langle m| \hat{\rho}|n \rangle \langle n|m
  \rangle =\langle n|\hat{\rho}|n \rangle .
\end{eqnarray}
Therefore, $P_{n}$ is nothing but the diagonal matrix element of density
operator in the Fock basis.

Let the system be in a pure state $|\psi \rangle$. Then
$\hat{\rho}=|\psi \rangle \langle \psi |$, and expanding $| \psi \rangle$ in
the series of Fock states $| n \rangle$,
\begin{eqnarray}
  |\psi \rangle =\sum_{m=0}^{\infty} C_{m} |m \rangle ,
\end{eqnarray}
we verify that $P_{n}$ is the squared modulo of coefficient $C_{n}$ in this
expansion:
\begin{eqnarray}
  P_{n}=|C_{n}|^{2}.
\end{eqnarray}

As a conclusion of this section consider the interpretation of quadrature
components in terms of physically measured values in the case of
electromagnetic field. In quantum electrodynamics usually the role of
coordinate and momentum of a field is played by vector potential and electric
field intensity, respectively. However, in quantum optical experiments the
vector potential is of no importance. That is why two values of electric field
intensity measured in the moments distanced by one--fourth of a period of
oscillations are taken as quadrature components. It can be easily shown that
one of these values is proportional to the vector potential. This question is
discussed in detail in \cite{200T}, and we will not describe it any
more.
\newpage

\setcounter{equation}{0}
\def\theequation{5.\arabic{equation}}

\begin{center}
{\Large 5. QUADRATURE COMPONENTS.\\
  MULTIMODE CASE}
\end{center}

In this section we generalize to the multi--dimensional case the
definitions and formulas of the previous one. Here we start with the
Hamiltonian of bosonic field typical for the field theory:
\begin{eqnarray}
  \hat{H}=\sum_{i=-\infty}^{\infty}\omega_{i} \left( \hat{a}_{i}^{+}
  \hat{a}_{i}+ \frac{1}{2} \right)
\end{eqnarray}
(we assume the Planck constant $\hbar$ be equal to unity). Thus, the field,
concrete nature of which is of no importance here, is represented by the
infinite number of noninteracting harmonic oscillators, numbered by index $i$
in expression (5.1). In (5.1) $\omega_{i}$ is the frequency of
$i$--th oscillator, $\hat{a}_{i}$ and $\hat{a}_{i}^{+}$ are boson
annihilation and creation operators for $i$--th mode of a field. They obey
the commutation rule:
\begin{eqnarray}
 \lefteqn{}
  &&\left[ \hat{a}_{i}, \hat{a}_{j}^{+} \right] = \delta_{ij}, \nonumber\\
  &&\left[ \hat{a}_{i}, \hat{a}_{j} \right]=
    \left[ \hat{a}_{i}^{+}, \hat{a}_{j}^{+} \right]=0.
\end{eqnarray}
Suppose, that somehow we cancel oscillations in all modes except the
finite number of them. Hamiltonian of such a field is
\begin{eqnarray}
  \hat{H}=\sum_{i=1}^{N} \omega_{i}\left( \hat{a}_{i}^{+} \hat{a}_{i}
  +\frac{1}{2} \right).
\end{eqnarray}
In this section we will consider so--called $N$--mode field with
Hamiltonian defined by (5.3). By definition introduce the quadrature
components. There will be $2N$ of them, i.e. the pair of canonically
conjugated variables per each mode:
\begin{eqnarray}
  \hat{x}_{i}=\frac{\hat{a}_{i}+\hat{a}_{i}^{+}}{\sqrt{2 \omega_{i}}},
  \hspace{0.4cm}
  \hat{p}_{i}=\sqrt{\frac{\omega_{i}}{2}}\hspace{0.15cm}
  \frac{\hat{a}_{i}-\hat{a}_{i}^{+}}{i}.
\end{eqnarray}
{}From (5.2) and (5.4) the commutation relations for quadrature
components follow:
\begin{eqnarray}
  \lefteqn{}
    &&\left[ \hat{x}_{i}, \hat{p}_{j} \right] =i \delta_{ij}, \nonumber\\
    &&\left[ \hat{x}_{i}, \hat{x}_{j} \right] =
      \left[ \hat{p}_{i}, \hat{p}_{j} \right] =0.
\end{eqnarray}
Hamiltonian (5.3) rewritten in terms of operators $\hat{x}_{i}$,
$\hat{p}_{i}$ has the usual oscillatory form:
\begin{eqnarray}
  \hat{H}= \sum_{i=1}^{N} \left( \frac{\hat{p}_{i}^{2}}{2}+
  \frac{\omega_{i}^{2} \hat{x}_{i}^{2}}{2} \right).
\end{eqnarray}
In analogy with section 4, define the mean values (moments) of quadrature
components. The moments of the first order are:
\begin{eqnarray}
  \lefteqn{}
    &&\langle \hat{x}_{i} \rangle= \mbox{Tr}
    \left(\hat{\rho}\hat{x}_{i}\right), \nonumber\\
    &&\langle \hat{p}_{i} \rangle = \mbox{Tr}
    \left( \hat{\rho} \hat{p}_{i} \right),
\end{eqnarray}
where $\hat{\rho}$ -- density operator describing $N$--mode quantized field.

Moments of the second order (dispersions) are defined as:
\begin{eqnarray}
  \lefteqn{}
    &&\sigma_{x_{i}x_{j}}= \langle \hat{x}_{i} \hat{x}_{j} \rangle -
      \langle \hat{x}_{i} \rangle \langle \hat{x}_{j} \rangle =
      \mbox{Tr} \left( \hat{x}_{i} \hat{x}_{j} \hat{\rho} \right)-
      \langle \hat{x}_{i} \rangle \langle \hat{x}_{j} \rangle, \nonumber\\
    &&\sigma_{p_{i}p_{j}}= \langle \hat{p}_{i} \hat{p}_{j} \rangle -
      \langle \hat{p}_{i} \rangle \langle \hat{p}_{j} \rangle =
      \mbox{Tr} \left( \hat{p}_{i} \hat{p}_{j} \hat{\rho} \right)-
      \langle \hat{p}_{i} \rangle \langle \hat{p}_{j} \rangle, \nonumber\\
    &&\sigma_{p_{i}x_{j}}=\sigma_{x_{j}p_{i}}=\frac{1}{2}\mbox{Tr}
      ~\left[ \left(\hat{x}_{j} \hat{p}_{i} + \hat{p}_{i} \hat{x}_{j} \right)
      \hat{\rho} \right] -
      \langle \hat{p}_{i} \rangle \langle \hat{x}_{j} \rangle.
\end{eqnarray}

To simplify the following formulas introduce the $N$--dimensional vectors:
\begin{eqnarray}
  \lefteqn{}
    &&{\bf n}=(n_{1},n_{2},\ldots n_{N}), \nonumber\\
    &&{\bf \alpha}=(\alpha_{1},\alpha_{2},\ldots \alpha_{N}).
\end{eqnarray}

Introduce the basis of Fock states for $N$--mode field. They are numbered by
$N$ integer parameters. The vacuum state is defined by formulas:
\begin{eqnarray}
  \hat{a}_{i}|{\bf 0} \rangle =0, \hspace{0.5cm} i=1,2,\ldots N.
\end{eqnarray}
States with fixed number of quanta in each mode are generated from vacuum
(5.9) by the action of creation operators $\hat{a}_{i}^{+}$:
\begin{eqnarray}
  |{\bf n} \rangle = \frac{\hat{a}_{1}^{+n_{1}}
  \hat{a}_{2}^{+n_{2}}\cdots \hat{a}_{N}^{+n_{N}}}{\sqrt{n_{1}! n_{2}!
  \cdots n_{N}!}} |{\bf 0} \rangle .
\end{eqnarray}
Fock states form the complete
orthonormalized system:
\begin{eqnarray}
\langle {\bf n} | {\bf m} \rangle
  =\delta_{n_{1} m_{1}} \delta_{n_{2} m_{2}}\cdots \delta_{n_{N}m_{N}}.
\end{eqnarray}
In analogy with (4.13) introduce the basis of coherent states
$|{\bf \alpha} \rangle$:
\begin{eqnarray}
  \hat{a}_{i}|{\bf \alpha} \rangle =
  \alpha_{i}|{\bf \alpha} \rangle .
\end{eqnarray}
In this case coherent states are parametrized by $N$ continuous complex
parameters $\alpha_{i}$, $i=1,2,\ldots N$.

The analogue of formula (4.14), expanding the coherent state in the
series of Fock states, for $N$--mode case is:
\begin{eqnarray}
  |{\bf \alpha} \rangle = \mbox{exp} \left[-\frac{|\alpha_{1}|^{2}
  +|\alpha_{2}|^{2}+\cdots +|\alpha_{N}|^{2}}{2} \right]
  \sum_{{\bf n}=0}^{\infty }
  \frac{\alpha_{1}^{n_{1}} \alpha_{2}^{n_{2}}\cdots
  \alpha_{N}^{n_{N}}}{\sqrt{n_{1}! n_{2}! \cdots n_{N}!}}
  |{\bf n} \rangle .
\end{eqnarray}

$N$--dimensional distribution function $P_{\bf n}$ is introduced
according to the definition:
\begin{eqnarray}
  P_{\bf n}
  = \mbox{Tr} ~\left[\hat{\rho} | {\bf n} \rangle
  \langle {\bf n}| \right]=
  \langle {\bf n} | \hat{\rho} | {\bf n} \rangle .
\end{eqnarray}


\setcounter{equation}{0}
\def\theequation{6.\arabic{equation}}

\begin{center}
{\Large 6. WIGNER QUASIDISTRIBUTION FUNCTION}
\end{center}

Quantum mechanics in its usual formulation deals with operators acting in
Hilbert spaces of states. However, there exist the alternative formulations
of quantum mechanics useful in considering concrete problems. An example of
such alternative approach is the formalism of Wigner function and Weyl
transformation.

Let quantum system be in a state described in terms of the density operator
$\hat{\rho}$. We consider the Hermitean operator $\hat{A}$ corresponding to
some physical quantity. In Weyl formulation of quantum mechanics the objects
corresponding to physical quantities are Weyl transformations, and those
corresponding to the states of a system are Wigner functions. By definition,
introduce the Weyl transformation of operator $\hat{A}$:
\begin{eqnarray}
  a({\bf p},{\bf q})
  = \int d{\bf u} ~e^{i{\bf qu}/\hbar}
  \langle \left. {\bf p} +\frac{\bf u}{2} \right|
  \hat{A} \left| {\bf p}- \frac{\bf u}{2} \right. \rangle ,
\end{eqnarray}
where $|{\bf p} \rangle$ is the eigenvector of momentum operator with
eigenvalue equal to ${\bf p}$. Another, equivalent representation reads
\begin{eqnarray}
  a({\bf p}, {\bf q})
  =\int d{\bf v} ~e^{i{\bf pv}/\hbar}
  \langle \left. {\bf q}-\frac{\bf v}{2} \right|
  \hat{A} \left| {\bf q}+ \frac{\bf v}{2} \right. \rangle ,
\end{eqnarray}
where $|{\bf q} \rangle$ is eigenvector of position operator with
eigenvalue equal to ${\bf q}$. We consider $N$--dimensional case,
so ${\bf q}$ and ${\bf p}$ are $N$--dimensional vectors.
{}From (6.1) and (6.2) it follows that the Weyl transformation of Hermitean
operator is real.

Operator $\hat{A}$ can be reconstructed from its Weyl transformation in the
following way:
\begin{eqnarray}
  \hat{A}=\frac{1}{(2\pi \hbar)^{N}}\int d{\bf p} ~d{\bf q}
  ~a({\bf p},{\bf q}) \hat{\Delta}({\bf p},
  {\bf q}),
\end{eqnarray}
where Hermitean operator $\hat{\Delta} ({\bf p},{\bf q})$ is
\begin{eqnarray}
  \hat{\Delta}({\bf p},{\bf q})= \int d{\bf u}
  ~e^{i{\bf qu}/\hbar} \left| {\bf p}
  - \frac{\bf u}{2} \right. \rangle \langle \left.
  {\bf p} + \frac{\bf u}{2} \right|, \nonumber\\
\end{eqnarray}
or
\begin{eqnarray}
  \hat{\Delta}({\bf p},{\bf q})= \int d{\bf v}
  ~e^{i{\bf pv}/\hbar} \left| {\bf q}
  + \frac{\bf v}{2} \rangle \langle {\bf q}
  - \frac{\bf v}{2} \right| .
\end{eqnarray}
The Weyl transformation can be also written as a trace including operator
$\hat{\Delta}$:
\begin{eqnarray}
  a({\bf p},{\bf q})=\mbox{Tr} ~\left[ \hat{A} \hat{\Delta}
  ({\bf p},{\bf q})\right].
\end{eqnarray}
The Wigner function is introduced as a Weyl transformation of the density
operator $\hat{\rho}$:
\begin{eqnarray}
  \lefteqn{}
    W({\bf p},{\bf q})
      &=&\int d{\bf v} ~e^{i{\bf pv}/\hbar}
      \langle \left. {\bf q}- \frac{\bf v}{2}
      \right| \hat{\rho} \left| {\bf q}
      + \frac{\bf v}{2} \right. \rangle
     \nonumber\\
   \lefteqn{}
    &=&\int d{\bf u} ~e^{i{\bf qu}/\hbar} \langle \left.
    {\bf p}+ \frac{\bf u}{2} \right| \hat{\rho} \left|
    {\bf p} - \frac{\bf u}{2} \right. \rangle.
\end{eqnarray}
It is real and normalized:
\begin{eqnarray}
  \int \frac{d{\bf p} ~d{\bf q}}{(2 \pi \hbar)^{N}}
  W({\bf p},{\bf q})=1.
\end{eqnarray}
However, it is not always positively definite. That is why it
cannot be interpreted as a distribution function in the phase space,
therefore it is called the quasidistribution function.

The mean value of operator $\hat{A}$ is obtained as integral of its Weyl
transformation $a({\bf p},{\bf q})$ and Wigner function
$W({\bf p},{\bf q})$:
\begin{eqnarray}
   \langle \hat{A} \rangle = \int W({\bf p},{\bf q})
   ~a({\bf p},{\bf q})
   \frac{d{\bf p} ~d{\bf q}}{(2 \pi \hbar)^{N}}  .
\end{eqnarray}

We will not describe the properties of Weyl transformation and Wigner
function, as well as those of some other quasidistribution functions
contiguous to $W({\bf p},{\bf q})$ refering the reader
to the monograph \cite{deGroot} and the review paper \cite{Scully}.

The formalism described above turns out to be very useful in considering
the photon statistics. Its main advantage is that the Wigner function
corresponding to the most general squeezed and correlated state of light
is an exponent of quadratic form with respect to quadrature components.
Corresponding integrals can be easily calculated giving the opportunity
to find explicitly characteristics of a system.


\setcounter{equation}{0}
\def\theequation{7.\arabic{equation}}

\begin{center}
{\Large 7. PHOTON DISTRIBUTION FUNCTIONS}
\end{center}

The photon statistics of nonclassical light  described by the
gaussian Wigner function has some interesting properties related
to the squeezing in one of the quadrature components \cite{Stol,Walls}.
The problem of finding the photon distribution function (PDF) for
one-mode gaussian states of electromagnetic field was
considered in Refs.\cite{Stol,Yuen,Walls},\cite{Vourdas}-\cite{[9]}.
The simplest expressions in terms
of Hermite polynomials of two variables with equal indices
were obtained recently in Ref.\cite{Ola1}, where the most general mixed
state of one--mode light described by a generic gaussian Wigner function
was treated.
The photon statistics of some special cases of gaussian pure states
for two--mode light has been studied in Refs. \cite{Car,sch}.

The aim of the present section is to discuss the results of
Ref.\cite{Ola1} in which the photon distribution for one--mode mixed
gaussian light was obtained explicitly, and the results of
Ref.\cite{Ola2} where the approach developed in \cite{Ola1} was
applied to the polymode case. All the information about the polymode
squeezing, mode correlations and thermal noise is contained for the
generic mixed state of the gaussian light in the quadrature means
and matrix elements of the real symmetric dispersion matrix
which determine the generic gaussian Wigner function. For $N$--mode
light the number of real parameters determining the gaussian
Wigner function is equal to $2N^{2}+3N$.
We show that the PDF of the $N$--mode
field state described by a generic gaussian Wigner function can be
expressed in terms of the Hermite polynomial of $2N$ variables with
equal pairs of indices.

It should be noted that the particle distribution
function and expression of density matrix nondiagonal elements
for polymode systems in terms of multivariable Hermite
polynomials were given for some gaussian states in Ref.\cite{[15]}.
However, the physical parameters used in this work were just the
parameters determining the inhomogeneous symplectic
canonical transformation of photon quadratures, which are related to
gaussian Wigner function parameters through a complicated
functional dependence.

The most general mixed squeezed state of the $N$--mode light with a
gaussian density operator $\hat{\varrho}$ is described by the Wigner
function $W({\bf p},{\bf q})$ of the generic gaussian form
(see, for example, \cite{DodMan89})
\begin{eqnarray}
  W({\bf p},{\bf q})
  =(\det {\bf M})^{-1/2}\exp\left
  [-\frac 12 \left({\bf Q}-\langle{\bf Q}\rangle)
  {\bf M}^{-1}({\bf Q}-
  \langle {\bf Q} \rangle \right) \right].
\end{eqnarray}

\noindent Here $2N$--dimensional vector
${\bf Q}=({\bf p},{\bf q})$ consists of $N$
components $p_1,\ldots ,p_N$ and $N$ components $q_1,\ldots ,q_N$,
operators $\hat {{\bf p}}$ and $\hat {{\bf q}}$ being the
quadrature components of photon creation
$\hat {{\bf a}}^{+}$
and annihilation $\hat {{\bf a}}$
operators (we use dimensionless variables and
assume $\hbar =1$):
\[
  \hat {{\bf p}}=\frac {\hat {{\bf a}}
  -\hat {{\bf a}}^{+}}{i\sqrt {2}},\qquad\hat {{\bf q}}
  =\frac {\hat {{\bf a}}+\hat {{\bf a}}^{+}}{\sqrt {2}}.
\]
\noindent$2N$ parameters $\langle p_i \rangle$ and
$\langle q_i \rangle$, $i=1,2,\ldots ,N$, combined into vector
$\langle {\bf Q}{\bf \rangle}$, are the average values
of quadratures,

\[
  \langle {\bf p} \rangle=\mbox{Tr}~(\hat{\varrho}
  \hat {{\bf p}}),\qquad
  \langle {\bf q} \rangle=\mbox{Tr}~(\hat{\varrho}
  \hat {{\bf q}}),
\]

\noindent A real symmetric dispersion matrix ${\bf M}$ consists
of $2N^2+N$ variances

\[
  {\cal M}_{\alpha\beta}=\frac 12\left\langle\hat Q_{\alpha}\hat
  Q_{\beta}+\hat Q_{\beta}\hat Q_{\alpha}\right\rangle -\left\langle
  \hat Q_{\alpha}\right\rangle\left\langle\hat Q_{\beta}\right\rangle
  ,\qquad\alpha ,\beta =1,2,\ldots ,2N.
\]

\noindent They obey certain constraints, which are nothing but the
generalized uncertainty relations \cite{DodMan89}.

Of course, one may choose different representations of the
statistical operator, for instance, coordinate representation or
various modifications of the coherent state representation (like
``normal'', ``antinormal'', ``positive'', etc.). However, the Wigner
function (7.1) seems the most suitable one, since for gaussian states
it is real and positive. Moreover, all coefficients of the quadratic
form in the exponent are real, and they have lucid physical meaning
(see, also ref.\cite{Scully}).

The photon distribution function is nothing but the probability  to
have $n_1$ photons in the first mode, $n_2$ photons in the second
mode, and so on. To point out that we discuss namely the photon
distribution we shall designate it in this section by symbol
${\cal P}_{\bf n}$, where vector
${\bf n}$ consists of $N$ nonnegative
integers:  ${\bf n}=(n_1,n_2,\ldots ,n_N)$.  This
probability is given by formula

\begin{eqnarray}
  {\cal P}_{\bf n}=\mbox{Tr}~\hat{\varrho }|{\bf n}
  \rangle \langle {\bf n}|,
    \nonumber
\end{eqnarray}
where $\hat{\varrho}$ is the density operator of the system of photons
under study, and the state $|{\bf n} \rangle$ is a common eigenstate
of the set of photon number operators $\hat {a}_i^{+}\hat {a}_i$,
$i=1,2,\ldots,N$:

\[
  \mbox{$\hat {a}_i^{+}\hat {a}_i$}{\bf |n} \rangle {\bf =}
  n_i{\bf |n} \rangle .
\]

\noindent The simplest way to find ${\cal P}_{\bf n}$
is to calculate the generating function for matrix elements
$\varrho_{\bf mn}$ of the density operator
$\hat{\varrho}$ in the Fock basis. This generating function, in turn,
is nothing but the matrix element of density operator in the coherent
state basis,

\begin{eqnarray}
  \langle \beta |\hat{\varrho }|\alpha \rangle =\exp
  \left( -\frac {|\alpha |^2}{2}-\frac {|\beta |^2}{2} \right)
  \sum_{{\bf m},{\bf n}={\bf 0}}^{\infty}\frac {(\beta^{*})^{\bf m}
  \alpha^{\bf n}}{({\bf m}!{\bf n}!)^{\frac 12}}
  \varrho_{\bf mn}.
  \label{2}
\end{eqnarray}
\noindent Hereafter $\alpha$ and $\beta$ without indices mean
$N$--dimensional vectors with complex components.

To proceed from the Wigner function to the matrix element of the
density operator in coherent state representation one should
calculate $2N$--dimensional overlap integral \cite{Scully}

\begin{eqnarray}
  \langle \beta |\hat{\varrho }|\alpha \rangle =
  \frac 1{(2\pi
  )^N} \int W({\bf p},{\bf q})W_{\alpha\beta}
  ({\bf p},{\bf q})~d{\bf p}~d{\bf q}
  ,\label{3}
\end{eqnarray}

\noindent where $W_{\alpha\beta}({\bf p},{\bf q})$
is the Wigner--Weyl transform of the operator
$|\alpha \rangle \langle \beta |$ which was found in
\cite{DodMan89},

\begin{eqnarray}
  W_{\alpha\beta}({\bf p},{\bf q})=
  &&2^N\exp\left[-\frac {|\alpha |^
  2}2-\frac {|\beta |^2}2-\alpha\beta^{*}-{\bf p}^2
  -{\bf q}^2
  \right. \nonumber\\
  &&\left. +\sqrt {2}\alpha ({\bf q}-i{\bf p})+
  \sqrt {2}\beta^{*}({\bf q}+i{\bf p})\right].
  \nonumber
\end{eqnarray}

\noindent Let us introduce $2N$--dimensional complex vector
\[
  \gamma =(\beta^{*},\alpha ),
\]
\noindent which is composed of two $N$--dimensional vectors, and
the $2N$--dimensional unitary matrix

\[
  {\bf U}=\frac 1{\sqrt {2}}\left(\begin{array}{cc}
  -i{\bf I}_N&i{\bf I}_N\\
  {\bf I}_N&{\bf I}_N\end{array}
  \right),
\]

\noindent satisfying relations

\begin{eqnarray}
  {\bf U}^{+}{\bf U}={\bf U}^T{\bf U}^{*}
  ={\bf I}_{2N},\qquad {\bf U}^{+}{\bf U}^{*}=
  {\bf U}^T{\bf U}=\left(
    \begin{array}{cc}
        0&{\bf I}_N\\
        {\bf I}_N&0
    \end{array}
  \right).\label{4}
\end{eqnarray}

\noindent(``$^{+}$'' means
the Hermitian conjugation, ``$^{*}$'' --- complex
conjugation, and ``$^{T}$'' --- transposition; ${\bf I}_N$ is
the $ N \times N $ identity matrix.)  Calculating the gaussian integral
on the right--hand side of eq.(7.3) we arrive at the expression

\begin{equation}
  \langle \beta |\hat{\varrho }|\alpha \rangle ={\cal
  P}_0\exp\left (-\frac {|\gamma |^2}2-\frac 12\gamma {\bf R}\gamma
  +\gamma {\bf Ry}\right),\label{5}
\end{equation}

\noindent where the symmetric $2N$--dimensional matrix ${\bf R}$
and $2N$--dimensional vector ${\bf y}$ are given by the relations
(we have taken into account the identities from eq.(7.4)):

\begin{equation}
  {\bf R}={\bf U}^{+}\left({\bf I}_{2N}
  -2{\bf M}\right) \left({\bf I}_{2N}
  +2{\bf M}\right)^{-1}{\bf U}^{*},
  \label{6}
\end{equation}

\begin{equation}
  {\bf y}
  =2{\bf U}^T({\bf I}_{2N}
  -2{\bf M})^{-1}\langle {\bf Q}
  \rangle .\label{7}
\end{equation}

\noindent\ Factor ${\cal P}_0$, which does not depend on vector
$\gamma$, is
nothing but the probability to register no photons.  It equals
\[
  {\cal P}_0=\left[\det\left({\bf M}
  +\frac 12{\bf I}_{2N}\right)\right
  ]^{-\frac 12}\exp\left[- \langle {\bf Q} \rangle \left(
  2{\bf M}+{\bf I}_{2N}\right )^{-1}
  \langle {\bf Q} \rangle \right].
\]

\noindent If $ \langle {\bf Q} \rangle ={\bf 0}$,
the probability to have no photons depends on
$2N-1$ parameters coinciding up to numerical factors with
coefficients of the characteristic polynomial of the variance
matrix.

Function
$\exp\left(|\gamma |^2/2\right)\langle \beta |\hat{\varrho }
|\alpha \rangle $
is the generating function for the
elements of density matrix in the photon number state basis.
Comparing eqs.(7.2) and (7.5) with the generating function for the
Hermite polynomials of $2N$ variables \cite{[14]},

\begin{equation}
  \exp\left(-\frac 12\gamma {\bf R}\gamma +\gamma {\bf R}
  {\bf y}\right)=\sum_{{\bf m},{\bf n}
  ={\bf 0}}^{\infty}\frac {\beta^{
  *{\bf m}}\alpha^{\bf n}}{{\bf m}!{\bf n}!}
  H_{{\bf m},{\bf n}}^{
  \{{\bf R}\}}({\bf y}),\label{8}
\end{equation}

\noindent we see that the photon distribution function
${\cal P}_{\hbox {\bf n}}$ can be
expressed through the ``diagonal'' multidimensional Hermite
polynomials:

\begin{equation}
  {\cal P}_{\bf n}={\cal P}_0\frac {H_{\bf nn}^{
  \{{\bf R}\}}({\bf y})}{{\bf n}!}.\label{9}
\end{equation}

\noindent This formula is a special case of the matrix element of
density operator in the Fock basis, which was obtained in
\cite{[15]} by the canonical transform method.
It may be shown that the multivariable Hermite polynomial is an
even function if the sum of its indices is an even number and the
polynomial with an odd sum of indices is an odd function.
This property is the natural generalization of the parity
properties of usual Hermite polynomials according to
which the energy states of harmonic oscillator for even excited
levels are even and for odd excited levels are odd ones.
Due to this parity property of polydimensional Hermite
polynomials the ``diagonal''
multivariable Hermite polynomial is an even function since
the sum of its indices is always an even number. Consequently, the
above photon distribution function is an even function.
In the limit case, when the only nonzero elements of the
quadrature dispersion matrix are the diagonal elements and they
are equal to $1/2$, the obtained distribution function (7.9)
becomes the product of $N$ one--dimensional Poisson distributions
describing the independent light modes in coherent states.

Let us discuss photon distributions for pure polymode states. The
formulas derived before hold for any gaussian state, which is, in
general, a mixed quantum state, i.e.
\begin{equation}
  \label{eq.10}
  \mu=\mbox{Tr} ~(\hat\varrho^2)\le 1,~~~~\mbox{Tr} ~\hat\varrho=1.
\end{equation}
\noindent In terms of Wigner function the ``mixing coefficient''
$\mu$ can be expressed as follows \cite{Scully,167}
  $$\mu=\frac{1}{(2\pi)^N}\int W^2({\bf p},{\bf q})
  ~d{\bf p}~d{\bf q}.$$
\noindent Evaluating this integral for the gaussian state we
arrive at a relation \cite{167,Ola2}
  $$\mu=2^{-N}(\det {\bf M})^{-1/2}.$$
\noindent Consequently, for the gaussian state restriction (7.10)
is equivalent to the inequality
\begin{equation}
  \label{eq.11}
  \det {\bf M}\ge(1/4)^N,
\end{equation}
\noindent which is nothing but one of the simplest forms of the
generalized uncertainty relations \cite{DodMan89}. In particular,
for $N=1$ inequality (7.11) turns into the Schr\"odinger--Robertson
uncertainty relation \cite{Sch30,Rob30}:

\[
  \sigma_{pp}\sigma_{qq}-\sigma_{pq}^2\ge\frac {1}{4}.
\]

Here we consider the pure gaussian states, when eqs.(7.10)
and (7.11) become strict equalities.  In this case the quantum
state is described in fact by a wave function, which has twice
less variables than the density matrix.  As a consequence, the
formulas for the PDF can be significantly simplified. Namely,
instead of polynomials of $2N$ variables one can manage with
the $N$--dimensional Hermite polynomials.

Let us consider an inhomogeneous linear canonical
transformation of photon creation and annihilation operators,

\begin{equation}
  \left(
    \begin{array}{c}
      \hat {{\bf b}}\\
      \hat {{\bf b}}^{+}
    \end{array}
  \right)=\Omega\left(
    \begin{array}{c}
      \hat {{\bf a}}\\
      \hat {{\bf a}}^{+}
    \end{array}
  \right)+\left(
    \begin{array}{c}
      {\bf d}\\
      {\bf d}^{*}
    \end{array}
  \right),\qquad
  \Omega =\left(
    \begin{array}{cc}
      \zeta&\eta\\
      \eta^{*}&\zeta^{*}
    \end{array}
  \right),\label{12}
\end{equation}
where $\Omega$ is a symplectic $2N$x$2N$--matrix  consisting of
four $N$--dimensional complex square blocks, and $\hbox {\bf d}$
is a complex $N$--vector. Designating the eigenstates of operators
$\hat {{\bf a}}$ and $\hat {{\bf b}}$ as $|\alpha \rangle$
and $|\beta \rangle$, respectively, one can write the formula (for its
derivation see, e.g., Refs.\cite{DodMan89,[19]}):

\begin{equation}
  \exp\left(\frac {|\alpha |^2}2+\frac {|\beta |^2}
  2\right) \langle \alpha^{*}|\beta \rangle ={\cal F}_0(\beta
  )\exp\left[-\frac 12
  \alpha^{*}\zeta^{-1}\eta\alpha^{*}+\alpha^{*}\zeta^{-1}(\beta
  -{\bf d}
  )\right],\label{13}
\end{equation}
where
\begin{equation}
  {\cal F}_0(\beta )=(\det\zeta )^{-\frac 12}\exp\left
  [\frac 12\beta\eta^{*}\zeta^{-1}\beta +\beta ({\bf d}^{*}
  -\eta^{*}\zeta^{-1}{\bf d})
  +\frac 12{\bf d}\eta^{*}\zeta^{-1}{\bf d}
  -|{\bf d}|^2\right].\label{14}
\end{equation}

Here $\beta$ should
be considered as a label of a state, while $\alpha^{*}$ as a
variable.  Expanding the right--hand side of (7.13) in a power series
of $\alpha^{*}$ with
account of (7.8), we obtain for the PDF in the state $|\beta \rangle$
(which is, in general, a squeezed coherent state) an expression
through the Hermite polynomial of $N$ variables:

\begin{equation}
  {\cal P}_{\bf n}=\frac {{\cal P}_0(\beta )}{{\bf n}
  !}\left|H_{\bf n}^{\{\zeta^{-1}\eta \}}\left(\eta^{-1}[\beta
  -{\bf d}
  ]\right)\right|^2,\qquad {\cal P}_0(\beta )=|{\cal F}_0(\beta )|^
  2.\label{15}
\end{equation}
A similar formula for the transition probabilities between the
initial and final energy states of a multimode parametric oscillator
was obtained in \cite{MalManTri73,DodMan89}.

Let us consider the PDF of the squeezed number state
$|{\bf m} \rangle$ defined by expansion

\[
  |\beta \rangle =\exp(-\frac {|\beta |^2}2)\sum_{{\bf m}={\bf 0}}^{
  \infty}\frac {(\beta )^{\bf m}}{({\bf m!})^{\frac 12}}
  |{\bf m}
  \rangle .
\]
Equations (7.13) and (7.14) lead to

\[
  {\cal P}_{\bf n}=|\det\zeta |^{-1}\exp\left[
  \mbox{Re}({\bf d}\eta^{*}\zeta^{-1}{\bf d})
  -2|{\bf d}|^2\right]
  \frac {\left|H_{\bf nm}^{\{{\bf R}\}}
  ({\bf L})\right|^2}{{\bf n}!{\bf m}!}.
\]
Here ${\bf m}$ is the label of the state, whereas
${\bf n}$ is a discrete vector variable.
$2N$x$2N$--matrix ${\bf R}$
and $2N$--vector ${\bf L}$ are expressed now in
terms of blocks of matrix $\Omega$ and vector ${\bf d}$
as follows,

\[
  {\bf R}=\left(
    \begin{array}{cc}
      \zeta^{-1}\eta & -\zeta^{-1}\\
      -\zeta^{-1}    & -\eta^{*}\zeta^{-1}
    \end{array}
  \right),\qquad {\bf L}={\bf R}^{*}\left(
    \begin{array}{c}
      -\zeta^{-1}{\bf d}\\
      {\bf d}^{*}-\eta^{*}\zeta^{-1}{\bf d}
    \end{array}
  \right).
\]
So it was demonstrated that in the case of polymode light in pure
gaussian state the expression for photon distribution function in
terms of Hermite polynomial with $2N$ indices (7.9) may be replaced
by the expression in terms of the Hermite
polynomial with $N$ indices.

In conclusion let us consider the particular case of one--mode mixed
light. The mixed squeezed state of light with density operator
$\hat \rho $ is described by the Wigner function $W(p,q)$ of the
generic gaussian form which contains five real parameters. Two
parameters are mean values of momentum
$\langle p \rangle$ and position $\langle q \rangle$ and other three
parameters are matrix elements of the real symmetric dispersion matrix
${\bf m}$
\begin {eqnarray}
  \lefteqn{}
    &&m_{11}=\sigma _{pp}
    =\mbox{Tr} \left( {\hat \varrho}{\hat p}^2 \right)
    -\langle p \rangle ^2,\nonumber\\
    &&m_{22}=\sigma _{qq}=\mbox{Tr} \left( {\hat
    \varrho}{\hat q}^2 \right) - \langle q \rangle ^2,\nonumber\\
    &&m_{12}=\sigma _{pq}=\frac{1}{2} \mbox{Tr} ~\left[
    {\hat \varrho}({\hat p\hat q}
    +{\hat q\hat p}) \right] -\langle p \rangle \langle q \rangle .
\end{eqnarray}

Below we will use invariant parameters

  $$T=\mbox{Tr} \hspace{0.15cm} {\bf m} =\sigma _{pp}
  +\sigma _{qq}$$
\noindent and
  $$d=\det {\bf m} =\sigma _{pp}\sigma _{qq}-\sigma _{pq}^2.$$

So the generic gaussian Wigner function for one--mode mixed light
has the form
\begin{eqnarray}
    W(p,q)=
    &&d^{-\frac{1}{2}}\exp
    \{-(2d)^{-1} \left[ \sigma_{qq}(p- \langle p \rangle )^{2}\right.
    \nonumber\\
    &&\left. +\sigma _{pp}(q- \langle q \rangle )^{2}
    -2\sigma _{pq}(p-
      \langle p \rangle )(q- \langle q \rangle )\right ]\}.
\end{eqnarray}

The photon distribution function of one--mode mixed light is
described by formulas (7.9), where matrix elements of the symmetric
matrix ${\bf R}$ are expressed
in terms of the dispersion matrix ${\bf m}$ as follows,
\begin {eqnarray}
  \lefteqn{}
    &&R_{11}= \left( T+2d+\frac{1}{2} \right) ^{-1}(\sigma _{pp}
    -\sigma _{qq}-2i\sigma _{pq}) =R_{22}^*,\nonumber\\
    &&R_{12}=\left( T+2d
     +\frac{1}{2} \right) ^{-1} \left( \frac{1}{2}-2d\right),
\end {eqnarray}
and arguments of Hermite polynomials are of the form
\begin{eqnarray}
  y_1={y_2}^*= \left( T-2d-\frac{1}{2} \right) ^{-1}
  \left[ (T-1) \langle z^* \rangle
  +(\sigma _{pp}-\sigma _{qq} +2i\sigma
  _{pq}) \langle z \rangle \right] .
\end{eqnarray}
The complex parameter $\langle z \rangle$ is given
by relation
\begin{eqnarray}
  \langle z \rangle =2^{-\frac{1}{2}} \left(
  \langle q \rangle +i \langle p \rangle \right) .
\label{defz}\end{eqnarray}
The probability to have no photons is given by formula
\begin{eqnarray}
  \lefteqn{}
    &&P_0= \left( d+\frac{1}{2}T+ \frac{1}{4} \right) ^{-\frac{1}{2}}
      \nonumber\\
    &&\times \exp \left[ \frac{- \langle p \rangle ^{2}(2\sigma _{qq}+
      1)- \langle q \rangle ^{2}(2\sigma _{pp}+1)
     +4\sigma _{pq} \langle p
      \rangle \langle q \rangle }{1+2T+4d} \right] .
\end{eqnarray}
Due to the physical meaning of dispersions, the parameters $\sigma _{pp}$
and $\sigma _{qq}$ must be nonnegative numbers, so the invariant
parameter $T$ is a positive number.  Also the determinant $d$ of the
dispersion matrix must be positive.

Thus we have shown that photon distribution functions for polymode mixed
gaussian light, for polymode squeezed number light  and for one--mode
mixed light may be expressed in terms of Hermite polynomials of several
variables. The physical meaning of the mixed gaussian state of light
may be understood if one takes into account that the pure multimode
gaussian state corresponds to the generalized correlated state introduced
by Sudarshan \cite{[20]}, who related these states to the symplectic
dynamical group. The one--mode correlated state was introduced in
\cite{Kur80} as the state minimizing Schr\"odinger uncertainty relation
\cite{Sch30,Rob30}. In Ref.\cite{[20]} the relation of this state to the
symplectic group $ISp(2,R)$ was clarified and the generalization
to multimode correlated state has been done on the basis
of the symmetry properties related to the group $ISp(2N,R)$.
As we have observed, the generalized correlated state constructed by
Sudarshan \cite{[20]} from the symmetry properties satisfies also the
condition of minimization of multimode Schr\"odinger uncertainty
relation, since the determinant of the dispersion matrix is equal to
the low limit of the uncertainty inequality just for these states.
{}From that point of view the multimode correlated states of light have
both properties of one--mode correlated state to give minimum of
uncertainties product and to be obtained by the full coset of the
linear canonical group by the stability group of the vacuum state.
The mixed gaussian states studied above may be considered as the
mixture of generalized correlated states plus thermal noise.
But this thermal noise may not be described by a single temperature
for all the modes. For a generic case of mixed gaussian light the
distinct temperatures may be prescribed to each mode. So the mixed
gaussian light is the generalized correlated light each normal
mode of which interacts with its own heat bath. The density operator
of such state may be obtained by a generic symplectic canonical
transform from the product of usual thermal states of the
electromagnetic field oscillators, each of them being described by
the Planck distribution formula with its own temperature.


\setcounter{equation}{0}
\def\theequation{8.\arabic{equation}}

\begin{center}
{\Large 8. OSCILLATIONS OF CUMULANTS IN\\ SLIGHTLY SQUEEZED STATES}
\end{center}

In this section we will reproduce the results of \cite{Pavel}
showing the cumulants and the ratio of cumulants to factorial moments
exhibit an oscillatory behaviour when one deals with slightly squeezed
states of light\footnote{Note that in \cite{Pavel} the misprint in
scale of coordinate axes for graphs of photon distribution function
$P_{n}$ appeared. The correct figures are drawn below in this
section.}.

Let us consider the most general mixed squeezed state of
one--mode light described
by the Wigner function $W(p,x)$ of the generic gaussian form with five
real parameters, $\langle x \rangle$, $\langle p \rangle$, $\sigma_{xx}$,
$\sigma_{pp}$, $\sigma_{px}$, specified in the previous section
(in this section we use letter $x$ instead of $q$ for the ``coordinate''
quadrature component, while $q$ is reserved for the order of cumulants
or factorial moments).
The generating function of the photon number distribution was
obtained  in \cite{Ola1}:
\begin{eqnarray}
    G(u)=P_{0}\left[\left(1-\frac{u}{\lambda_{1}}\right)
    \left(1-\frac{u}{\lambda_{2}}\right)\right]^{-1/2}
    \mbox{exp}\left[\frac{u\xi_{1}}{u-\lambda_{1}}+\frac{u\xi_{2}}
    {u-\lambda_{2}}\right] \hspace{0.15cm} \mbox{,}
\end{eqnarray}
where
\begin{eqnarray}
  \lefteqn{}
    &&\lambda_{1}=\left(\sqrt{R_{11}R_{22}}
    -R_{12}\right)^{-1} \hspace{0.15cm}
    \mbox{,}
    \hspace{1cm} \lambda_{2}=-\left(\sqrt{R_{11}R_{22}}
    +R_{12}\right)^{-1}
    \hspace{0.15cm} \mbox{,}\nonumber\\
    &&\xi_{1}=
    \frac{1}{4}\left(1-\frac{R_{12}}{\sqrt{R_{11}R_{22}}}\right)
    \left(y_{1}^{2}R_{11}
    +y_{2}^{2}R_{22}-2\sqrt{R_{11}R_{22}}y_{1}y_{2}
    \right) \hspace{0.15cm} \mbox{,} \nonumber\\
    &&\xi_{2}
    =\frac{1}{4}\left(1+\frac{R_{12}}{\sqrt{R_{11}R_{22}}}\right)
    \left(y_{1}^{2}R_{11}+y_{2}^{2}R_{22}
    +2\sqrt{R_{11}R_{22}}y_{1}y_{2}
    \right) \hspace{0.15cm} \mbox{.} \nonumber
\end{eqnarray}

It was already mentioned in introduction that the photon distribution
function exhibits an oscillatory behaviour if we deal with highly
squeezed states (\hspace{0.05cm} $T=\sigma_{pp}
+\sigma_{xx}\gg 1$ \hspace{0.05cm} )
for large values of the parameter $z$.
A question arises: is it possible to obtain a similar ``abnormal''
behaviour of other characteristics of the photon distribution, namely,
introduced in section 2
cumulants, factorial moments and their ratio $H_{q}$ ?
If yes, then in what region of parameters can such
anomalies reveal themselves?

The direct differentiation of function $\mbox{ln}~G(u)$ at $u=1$ yields
the cumulants (see section 2)
\begin{eqnarray}
    K_{q}=\frac{(q-1)!}{\langle n \rangle^{q}}\left[\frac{1}
    {(\lambda_{1}-1)^{q}}
    \left(\frac{1}{2}+q\frac{\xi_{1}\lambda_{1}}{1-\lambda_{1}}\right)
    +\frac{1}{(\lambda_{2}-1)^{q}}
    \left(\frac{1}{2}+q\frac{\xi_{2}\lambda_{2}}{1-\lambda_{2}}\right)
    \right],
\label{gencum}\end{eqnarray}
with the average number of photons $\langle n \rangle$  \cite{Ola1}
\begin{eqnarray}
    \langle n \rangle=\frac{T-1}{2}+|z|^{2} \hspace{0.15cm}
    \mbox{.} \nonumber
\end{eqnarray}
The Schr\"odinger--Robertson uncertainty relation results in
restrictions
\begin{eqnarray}
    T\ge 1, \qquad d\geq \frac{1}{4},
\end{eqnarray}
which, in turn, lead to inequalities
\begin{eqnarray}
  \lefteqn{}
  &&\lambda_{1}>1,\\
  &&\lambda_{2}<0 \hspace{1cm}\mbox{or} \hspace{1cm} \lambda_{2}>1
  \hspace{0.15cm} \mbox{.}
\end{eqnarray}
The expression in brackets in eq.(\ref{gencum}) consists of two terms:
\begin{eqnarray}
    \frac{1}{(\lambda_{1}-1)^{q}}
    \left(\frac{1}{2}+q\frac{\xi_{1}\lambda_{1}}{1-\lambda_{1}}\right)
    \hspace{0.15cm} \mbox{,}
\end{eqnarray}
\begin{eqnarray}
    \frac{1}{(\lambda_{2}-1)^{q}}
    \left(\frac{1}{2}+q\frac{\xi_{2}\lambda_{2}}{1-\lambda_{2}}\right)
     \hspace{0.15cm} \mbox{.}
\end{eqnarray}
The first term has constant sign. The second one oscillates in
the case $\lambda_{2}<0$. With the aim to obtain oscillations of
the whole function $K_{q}$ we will treat only this case,
\begin{eqnarray}
    \lambda_{2}<0 \hspace{0.15cm} \mbox{.} \nonumber
\end{eqnarray}
Then,
\begin{eqnarray}
    \frac{1}{(\lambda_{2}-1)^{q}}
    =\frac{(-1)^{q}}{(1+|\lambda_{2}|)^{q}} \hspace{0.15cm} \mbox{.}
    \nonumber
\end{eqnarray}
However, then it follows that
\begin{eqnarray}
    |\lambda_{2}|\geq \lambda_{1}>1 \hspace{0.15cm} \mbox{,}
   \nonumber
\end{eqnarray}
and the alternating term diminishes faster than the constant sign
term. Terms $1/|\lambda_{2}-1|$ and
$1/(\lambda_{1}-1)$ are most close to each other if
\begin{eqnarray}
    d=\frac{1}{4} \hspace{1cm} \mbox{(the pure state),} \nonumber
\end{eqnarray}
that is used in the following.

  First of all we consider the simplest case when value of
$z$ as given by (\ref{defz}) equals zero. Then
\begin{eqnarray}
  \lefteqn{}
    &&K_{q}=\frac{\beta^{q/2}}{\langle n \rangle ^{q}}(q-1)!
    T_{q}(\alpha) \hspace{0.15cm} \mbox{,}\nonumber\\
    &&F_{q}
    =\frac{\beta^{q/2}}{\langle n \rangle ^{q}}q!P_{q}(\alpha)
    \hspace{0.15cm} \mbox{,}
\end{eqnarray}
where
\begin{eqnarray}
    \beta=d+\frac{1}{4}
    -\frac{T}{2} \hspace{0.15cm} \mbox{,} \hspace{1cm}
    \alpha=\frac{T-1}{\sqrt{4d+1
    -2T}} \hspace{0.15cm} \mbox{,}\nonumber
\end{eqnarray}
$T_{q}(\alpha)$ and $P_{q}(\alpha)$ are the Chebyshev polynomials
of the first kind and Legendre polynomials, respectively.
Let us note that the arguments of the polynomials are purely
imaginary but the whole expressions for the moments are real,
surely.

For $H_{q}$ we obtain the expression:
\begin{eqnarray}
    H_{q}
    =\frac{T_{q}(\alpha)}{qP_{q}(\alpha)} \hspace{0.15cm} \mbox{.}
\end{eqnarray}
Taking $d=1/4$ we have:
\begin{eqnarray}
  \lefteqn{}
    &&K_{q}=\left(\frac{2}{1-T}\right)^{q/2}(q-1)!T_{q}(\alpha)
    \hspace{0.15cm} \mbox{,}
    \nonumber\\
    &&F_{q}=\left(\frac{2}{1-T}\right)^{q/2}q!P_{q}(\alpha)
    \hspace{0.15cm} \mbox{,}
    \nonumber\\
    &&\alpha=\sqrt{\frac{1-T}{2}} \hspace{0.15cm} \mbox{.} \nonumber
\end{eqnarray}

In this case the curve $H_{q}$ has step--like shape at
$(T-1)\rightarrow 0$; steps become smoothed as $T$ grows (fig.$1$).
We should note that direct limit $T\rightarrow 1$ shows the
discontinuous character of the function $H_{q}(T)$ at $T=1$.
The point is that at $T=1$ we are dealing with the usual Poisson
distribution (let us remind that we treat a case $d=1/4$,
$|z|=0$), where $H_{q}=\delta_{q1}$, i.e. $H_{1}=1$, $H_{q}=0$
at $q\neq 1$, which differs from the behaviour of $H_{q}$
at $(T-1)=10^{-5}$ depicted in fig.$1$.

Consider now the case $|z|\neq 0$. Since the photon distribution
function is invariant with respect to rotation in a phase space,
without loss of generality we can consider
$\sigma_{xx}=\sigma_{pp}$ ($\sigma_{px}\neq 0$ --
correlated state). By appropriate choice of the phase of $~z~$
($\langle x \rangle =-\langle p \rangle $) we
cancel the linearly increasing term
$q\xi_{1}\lambda_{1}/(1-\lambda_{1})$ in
(8.6).  Moreover, the analogous linear term
$q\xi_{2}\lambda_{2}/(1-\lambda_{2})$ in (8.7) becomes maximal at
fixed $|z|$. Thus we have left only two variable parameters $T$
and $|z|$, and formula (8.2) has the following final form:
\begin{eqnarray}
    K_{q}=\frac{(q-1)!}{\left(\frac{T-1}{2}
    +|z|^{2}\right)^{q}}
    \left[\frac{1}{2(\lambda_{1}-1)^{q}}
    +\frac{(-1)^{q}}{(1+|\lambda_{2}|)^{q}}\left(\frac{1}{2}
    -q\frac{\xi_{2}|\lambda_{2}|}{1+|\lambda_{2}|}\right)
    \right] \hspace{0.15cm} \mbox{,}
\end{eqnarray}
where
\begin{eqnarray}
   \lefteqn{}
     &&\lambda_{1}=-\lambda_{2}
      =\sqrt{\frac{T+1}{T-1}} \hspace{0.15cm} \mbox{,}
     \nonumber\\
     &&\xi_{2}=2\left(\frac{T}{\sqrt{T^{2}-1}}+1\right)|z|^{2}
     \hspace{0.15cm} \mbox{.}
     \nonumber
\end{eqnarray}

In the case of large $T$ (a highly squeezed state) we can obtain the
finite number of oscillations of $K_{q}$ considering large values of
$|z|$. However, the average number of photons in corresponding states
is large, and the amplitude of oscillations decreases exponentially
due to the factor $1/\langle n \rangle ^{q}$. Remind that in this very
case the strong oscillations of the photon distribution function can
be observed.

Now let us consider the case of the slightly squeezed state,
$y=(T-1)\ll 1$ when the photon distribution function does not
oscillate. Impose also an additional condition
\begin{eqnarray}
    \gamma=\frac{|z|^{2}}{\sqrt{y/2}}\gg 1
    \hspace{0.15cm} \mbox{,} \nonumber
\end{eqnarray}
that makes possible to obtain approximate formulas for the
functions $K_{q}$, $F_{q}$, and $H_{q}$.
For $K_{q}$, we have the following approximate expression:
\begin{eqnarray}
    K_{q}=q!(-1)^{q-1}\gamma^{1-q} \hspace{0.15cm} \mbox{.}
\end{eqnarray}
Then recursion relation (2.9) yields:
\begin{eqnarray}
    F_{q}=q!(-1)^{q}\gamma^{-q}L_{q}^{-1}(\gamma)
    \hspace{0.15cm} \mbox{,}
\end{eqnarray}
where $L_{q}^{-1}(x)$ are generalized Laguerre polynomials.
For $H_{q}$, with $ q \ll \gamma $ we have:
\begin{eqnarray}
    H_{q}=K_{q}/F_{q}
     =-\frac{\gamma}{L_{q}^{-1}(\gamma)}\approx (-1)^{q+1}
    q!\gamma^{1-q} \ll 1 \hspace{0.15cm} \mbox{.}
\end{eqnarray}
(If $\gamma\gg q$ the term with highest power of $\gamma$ dominates
over the rest of sum in $L_{q}^{-1}(\gamma)$, and
$F_{q} \rightarrow 1$ as for Poisson distribution). The exact shape
of the function $H_{q}$ is shown in fig.$2$.
The distribution function $P_{n}$ does not oscillate (fig.$2^{a}$).

However, the most abrupt oscillations of the functions $K_{q}$ and
$H_{q}$ have been obtained when $(T-1)\ll 1$, but condition
$\gamma\gg 1$ is not valid. The corresponding curves are shown in
figs.$3$, $3^{a}$. Note that the photon distribution function is
smooth again being approximately equal to zero at $q\neq 1$.

The most regular oscillating patterns of $K_{q}$ and $H_{q}$ are
seen at $(T-1)\sim 0.1$, $|z|\sim 1$ (figs.$4$, $4^{a}$).

The alternating sign cumulants are typical also for the fixed
multiplicity distribution, i.e. for $P_{n}=\delta_{nn_{0}}$
($n_{0}=const$) \cite{DreminPhysLett}.
Let us note that there exist smooth multiplicity distributions
which give rise to cumulants oscillating with larger period
(see, \cite{Dremin}).

Finally we consider the opposite case when the photon
distribution function
$P_{n}$ exhibits strong oscillations while $K_{q}$ and $H_{q}$
behave smoothly. Such a behaviour is typical at $T\sim 100$,
$|z|\sim 1$  when $K_{q}$ exponentially grows while
$H_{q}$ monotonically decreases with $q$ (fig.$5$).

Thus we have shown that the cumulants of the photon distribution
function for one--mode squeezed and correlated light at finite
temperature possess strongly oscillating behaviour in the region of
slight squeezing where the photon distribution function itself has
no oscillations. And vice versa in the region of large squeezing,
where the photon distribution function strongly oscillates, the
cumulants behave smoothly. Hence, the behaviour of cumulants may
provide a very sensitive method of detecting very small
squeezing and correlation phenomena due to the presence of strong
oscillations.

\newpage

\setcounter{equation}{0}
\def\theequation{9.\arabic{equation}}

\begin{center}
{\Large 9. PHOTON DISTRIBUTION FUNCTION AND ITS\\ MOMENTS
   FOR SCHR\"ODINGER CAT STATES}
\end{center}

As we said inintroduction, there exists another example of nonclassical
light state (Schr\"odinger cat state) considered for the first time in
\cite{DM74}. The paradox invented by Schr\"odinger is well--known:
let's consider a cat in two states; one corresponding to alive cat and
another --- to dead one, both described by their own wave functions.
If the wave function of alive cat is denoted by $\psi_{a}$, and that
of dead cat is denoted by $\psi_{d}$ (since cat is a macroscopic object,
both of these functions, of course, depend on a large number of
variables), the superposition principle of quantum mechanics enables
the existence of a cat in the states described by the wave functions
\begin{eqnarray}
  \lefteqn{}
    &&\psi_{1}=\frac{1}{\sqrt{2}}(\psi_{a}+\psi_{d}),
      \hspace{0.4cm} \mbox{and}\nonumber\\
    &&\psi_{2}=\frac{1}{\sqrt{2}}(\psi_{a}-\psi_{d}).
\end{eqnarray}
Certainly, the cat is too complicated object to be considered by
the quantum physics. Hence, we have to choose somewhat simpler object.

Consider one--mode electromagnetic field in coherent states
$|\alpha \rangle$ and $|- \alpha \rangle$ and form even and odd
normalized combinations of these states:
\begin{eqnarray}
  \lefteqn{}
    &&\left| \alpha_{+} \rangle \right. =N_{+}\left( |\alpha \rangle +
      |-\alpha \rangle \right) \hspace{0.4cm} \mbox{and} \nonumber\\
    &&\left| \alpha_{-} \rangle \right. =N_{-}\left( |\alpha \rangle -
      |-\alpha \rangle \right).
\end{eqnarray}
The normalization constants are given by relations:
\begin{eqnarray}
  N_{+}=\frac{\mbox{exp} \left( |\alpha|^{2}/2 \right)}
        {2\sqrt{\mbox{ch}|\alpha|^{2}}}, \hspace{0.5cm}
  N_{-}=\frac{\mbox{exp} \left( |\alpha|^{2}/2 \right)}
        {2\sqrt{\mbox{sh}|\alpha|^{2}}}.
\end{eqnarray}
The photon distribution function for such states can be easily
calculated using (4.14) \cite{Nikonov93}:
\begin{eqnarray}
  \lefteqn{}
    &&P_{n}^{(+)}
    =\frac{1}{\mbox{ch}|\alpha|^{2}} \frac{|\alpha|^{2n}}{n!}
      \left[ 1+(-1)^{n} \right],
    \nonumber\\
    &&P_{n}^{(-)}
      =\frac{1}{\mbox{sh}|\alpha|^{2}} \frac{|\alpha|^{2n}}{n!}
      \left[ 1-(-1)^{n} \right].
\end{eqnarray}
The multidimentional generalization of the above formulas is
straightforward \cite{Ansari94}.
So, fast oscillations of the photon distribution function for the
Schr\"odinger cat states are connected with the absence of states
with even numbers of photons in odd cat states and states with odd
numbers in even cat states.

Using eq.(2.2) one can find the generating functions for quantum
distributions in the even and odd cat states
\begin{eqnarray}
  G^{(+)}(z)=\frac{\mbox{ch} \left( |\alpha|^{2}z \right)}
           {\mbox{ch} ~|\alpha|^{2}},
  \hspace{0.5cm}
  G^{(-)}(z)=\frac{\mbox{sh} \left( |\alpha|^{2}z \right)}
           {\mbox{sh} ~|\alpha|^{2}}.
\end{eqnarray}

The average numbers of photons in these states read
\begin{eqnarray}
  \lefteqn{}
    &&\langle n \rangle ^{(+)}=|\alpha|^{2} \mbox{th} ~|\alpha|^{2}
      \hspace{0.4cm} \mbox{for even state,}
      \nonumber\\
    &&\langle n \rangle ^{(-)}
      =\frac{|\alpha|^{2}}{\mbox{th} ~|\alpha|^{2}}
      \hspace{0.4cm} \mbox{for odd state.}
\end{eqnarray}
Functions $G^{(+)}(z)$ and $G^{(-)}(z)$ can be easily differentiated
giving the explicit expressions for factorial moments:
\begin{eqnarray}
  \lefteqn{}
    &&F_{q}^{(+)}=\left( \mbox{th} ~|\alpha|^{2} \right) ^{-q},
      \hspace{0.4cm} q=2k, \nonumber\\
    &&F_{q}^{(+)}=\left( \mbox{th} ~|\alpha|^{2} \right) ^{-(q-1)},
      \hspace{0.4cm} q=2k+1,
\end{eqnarray}
and
\begin{eqnarray}
  \lefteqn{}
    &&F_{q}^{(-)}=\left( \mbox{th} ~|\alpha|^{2} \right) ^{q},
      \hspace{0.4cm} q=2k, \nonumber\\
    &&F_{q}^{(-)}=\left( \mbox{th} ~|\alpha|^{2} \right) ^{q-1},
      \hspace{0.4cm} q=2k+1.
\end{eqnarray}
The logarithm of $G^{(+)}(z)$ and $G^{(-)}(z)$ can not be so easily
differentiated, so to find the cumulants we have to solve numerically
the system of linear algebraic equations (2.9). As a result we have
cumulants oscillating with extremely rapidly growing amplitude in the
whole range of values of the parameter $|\alpha|$. The function
$H_{q}=K_{q}/F_{q}$ behaves in a similar way. To illustrate it we
plot here in fig.$6$ the graph of function $H_{q}$ for
odd cat state at $|\alpha|^{2}=2$.

\newpage

\setcounter{equation}{0}
\def\theequation{10.\arabic{equation}}

\begin{center}
{\Large 10. CONCLUSION}
\end{center}
New developments in studies of nonclassical states of the
electromagnetic field, from one side, and of multiplicity distributions
in high energy particle interactions, from the other side, provide
some hope for possible interrelation of the states under consideration.
The problem has matured for further exploration.

In this brief review we tried to demonstrate the methods applied for
description of nonclassical fields. No attempt was made to use them
directly in treatment of high energy multiparticle production even
though we consider it as a very promising problem to be studied later.
We can not resist to stress that the similarity  of Lagrangians of
quantum electrodynamics and quantum chromodynamics, not to say about
their obvious difference, could have much deeper meaning with
nonclassical fields playing an important role. Squeezed and correlated
gluonic fields could be as crucial as those of photons. Our hope for
success of the idea is based both on the similarity of theoretical
analyses of particle distributions in those cases and on
some experimental indications discussed above. In particular,
we have shown that the oscillatory behaviour of cumulants of the
multiplicity distributions predicted first in quantum chromodynamics
for high energy particle production processes reveals itself also for
the slightly squeezed states as well as for even and odd states
(Schr\"odinger cats). It could be a signature of some collective
effects being important in both cases.

We would consider our task fulfilled if the review helps somebody
to promote the analogy further.
\vspace{0.3cm}
\begin{center}
ACKNOWLEDGMENTS
\end{center}
This work is supported in part by Russian program "Fundamental
nuclear physics", by Russian fund for fundamental research, and by
JSPS.

\newpage

\begin{center}
    {\Large FIGURE CAPTIONS}
\end{center}
\renewcommand{\baselinestretch}{1}

Figure $1$: The behaviour of the function $H_{q}$ defined in (2.8) at
$d=1/4$, $|z|=0$; parameter $T$ is varied: (1) $(T-1)=10^{-5}$, (2) $T=1.1$,
(3) $T=1.2$.

Figure $2$: The behaviour of the function $H_{q}$ at $d=1/4$, $T=1.01$;
the curves (1) and (2) correspond to the values
$|z|^{2}=1.01, \hspace{0.1cm} 0.8$, respectively.

Figure $2^{a}$: The photon distribution function at $d=1/4$, $T=1.01$;
the curves in order of the lowering maxima correspond respectively to the
values $|z|^{2}=1.01, \hspace{0.1cm} 0.8$.

Figure $3$: The cumulants of the photon distribution function at $d=1/4$,
$(T-1)=10^{-5}$, $|z|^{2}=0.01$.

Figure $3^{a}$: The behaviour of the function $H_{q}$ at $d=1/4$,
$(T-1)=10^{-5}$; parameter $|z|$ is varied:
(1) $|z|^{2}=0.01$, (2) $|z|^{2}=0.005$, (3) $|z|^{2}=0.01$.

Figure $4$: The cumulants of the photon distribution function at $d=1/4$,
$T=1.1$, $|z|^{2}=1.1$.

Figure $4^{a}$: The behaviour of the function $H_{q}$ at $d=1/4$, $T=1.1$;
parameter $|z|$ is varied: (1) $|z|^{2}=2$, (2) $|z|^{2}=1.1$.

Figure $4^{b}$: The photon distribution function at $d=1/4$, $T=1.1$;
the curves in order of the lowering maxima correspond respectively to the
values $|z|^{2}=2, \hspace{0.1cm} 1.1$.

Figure $5$: The smooth curve for the function $H_{q}$ and the oscillating
photon distribution function $P_{n}$ at $d=1/4$, $T=100$, $|z|=1$.

Figure $6$: The behaviour of the function $H_{q}$ for odd Schr\"odinger cat
state at $|\alpha|^{2}=2$.


\newpage


\begin{thebibliography}{99}

\bibitem{Glauber63} R.J.Glauber, Phys.Rev.Lett. {\bf 10} (1963) 84.
\bibitem{SudarshanKla} J.R.Klauder and E.C.G.Sudarshan, Fundamentals
                       of Quantum Optics
                       (Benjamin, New York, 1970).
\bibitem{KlaSkag} J.R.Klauder and B.S.Skagerstam, Coherent States
                  (World Scientific, Singapore, 1985).
\bibitem{MalTri69} I.A.Malkin, V.I.Man'ko, and A.D.Trifonov, Phys. Lett.
                   A{\bf 30} (1969) 414.
\bibitem{MalMan70} I.A.Malkin and V.I.Man'ko, Phys. Lett A{\bf 32}
                   (1970) 243.
\bibitem{MalManTri70} I.A.Malkin, V.I.Man'ko, and D.A.Trifonov, Phys.
                      Rev. D{\bf 2} (1970) 1371.
\bibitem{MalManTri73} I.A.Malkin, V.I.Man'ko, and D.A.Trifonov,
                      J. Math. Phys. {\bf 14} (1973) 576.
\bibitem{Er1880} P.Ermakov, Univ. Izv. Kiev {\bf 20} N9 (1880) 1.
\bibitem{Lewis} H.R.Lewis, Phys. Rev. Lett. {\bf 18} (1967) 510.
\bibitem{MalJETP} I.A.Malkin and V.I.Man'ko, Zh. Eksp. Teor. Fiz.
                  {\bf 58} (1970) 721 [Sov. Phys. - JETP {\bf 31}
                  (1970) 386].
\bibitem{MalTriJETP} I.A.Malkin, V.I.Man'ko, and D.A.Trifonov,
                     Zh. Eksp. Teor. Fiz. {\bf 59} (1970) 1746
                     [Sov. Phys. - JETP {\bf 32} (1971) 949].
\bibitem{MalDod72} V.V.Dodonov, I.A.Malkin, and V.I.Man'ko,
                   Physica {\bf 59} (1972) 241.
\bibitem{MalMan79} I.A.Malkin and V.I.Man'ko, Dynamical Symmetries and
                   Coherent States of Quantum Systems (Nauka,
                   Moscow, 1979) [In Russian].
\bibitem{DodMan89} V.V.Dodonov and V.I.Man'ko, in
                   ``Invariants and Evolution of Nonstationary
                   Quantum Systems'', ed. M.A.Markov,
                   Proceedings of the Lebedev Physical
                   Institute, Vol.183, Nauka, Moscow, 1987
                   [translated by Nova Science, Commack, N.Y., 1989].
\bibitem{Cas} O.Casta\~nos, J.Fernandez--Nunez, and E.Martinez,
              Lett. Math. Phys. {\bf 23} (1991) 51.
\bibitem{Pro} G.Profilo and G.Soliani,
              Ann. Phys. (N. Y. ) {\bf 229} (1994) 160.
\bibitem{Castanes} O.Casta\~nos, R.Lopez--Pena, and V.I.Man'ko, J. Phys. A:
                   Math. Gen.  {\bf 27} (1994) 1751.
\bibitem{Heis27} W.Heisenberg, Z. Phys. {\bf 43} (1927) 172.
\bibitem{Stol} D.Stoler, Phys. Rev. D{\bf 1} (1970) 3217.
\bibitem{Mish} E.A.Mishkin, Bull. Am. Phys. Soc. {\bf 15} (1970) 89.
\bibitem{Yuen} H.P.Yuen, Phys. Rev. A{\bf 13} (1976) 2226.
\bibitem{Holl} J.N.Hollenhorst, Phys. Rev. D{\bf 19} (1979) 1669.
\bibitem{Walls} D.F.Walls, Nature {\bf 306} (1983) 141.
\bibitem{Kur80} V.V.Dodonov, E.V.Kurmyshev, and V.I.Man'ko,
                Phys. Lett. A{\bf 79} (1980) 150.
\bibitem{Sch30} E.Schr\"odinger, Ber. Kgl. Akad. Wiss. Berlin, {\bf 24}
                (1930) 296.
\bibitem{Rob30} H.P.Robertson, Phys. Rev. {\bf 35} (1930) 667.
\bibitem{DM74} V.V.Dodonov, I.A.Malkin, and V.I.Man'ko,
               Physica {\bf 72} (1974) 597.
\bibitem{Yur86} B.Yurke and D.Stoler, Phys. Rev. Lett. {\bf 57} (1986) 13.
\bibitem{WeinVourdas} A.Vourdas and R.M.Weiner, Phys. Rev. A{\bf 36}
                      (1987) 5866.
\bibitem{SchlW} W.Schleich and J.A.Wheeler, J. Opt. Soc. Am. B{\bf 4}
               (1987) 1715.
\bibitem{Rosa} V.V.Dodonov, O.V.Man'ko, V.I.Man'ko, and
               L.Rosa, Phys. Lett. A{\bf 185} (1994) 231.
\bibitem{Nikonov93} V.V.Dodonov, D.E.Nikonov, and V.I.Man'ko,
                    ``Even and Odd Coherent States (Schr\"odinger Cat States)
                    for Multimode Parametric Systems'', Preprint Univ. of
                     Naples INFN-NA-IV-93/49, DSF-T-93/49 (1993),
                     to appear in Phys. Rev. A.
\bibitem{Ola1} V.V.Dodonov, O.V.Man'ko, and V.I.Man'ko,
               Phys. Rev. A{\bf 49} (1994) 2993.
\bibitem{Ola2} V.V.Dodonov, O.V.Man'ko, and
               V.I.Man'ko, Phys. Rev. A{\bf 50}  (1994) 813.
\bibitem{Ansari94} N.A.Ansari and V.I.Man'ko, Phys. Rev. A{\bf 50}
                  (1994) 1942.
\bibitem{UA5}  UA5 Collab., G.J.Alner et al., Phys. Rep. {\bf 154} (1987) 247;
               R.E.Ansorge et al., Z. Phys. C{\bf 43} (1989) 357.
\bibitem{DELPHI} DELPHI Collab., P.Abreu et al., Z. Phys. C{\bf 52} (1991) 271.
\bibitem{OPAL} OPAL Collab., P.D. Acton et al., Z. Phys. C{\bf 58} (1993) 387.
\bibitem{Botke} J.C.Botke, D.J.Scalapino, and R.L.Sugar, Phys. Rev. D{\bf 9}
                (1974) 813; D{\bf 10} (1974) 1604.
\bibitem{Art} M.Artoni and J.L.Birman, Phys. Rev. B{\bf 44} (1991) 3736;
              in Proceedings of the Third International Workshop on
              Squeezed States and Uncertainty Relations (Baltimore, 10-14
              August 1993) eds. D.Han, Y.S.Kim, M.H.Rubin, Y.-H.Shih, and
              W.W.Zachary, NASA Conference Publication No.3270 (1994), p.495.
\bibitem{Pavlov} S.T.Pavlov and A.V.Prokhorov, Fiz. Tv. Tela, {\bf 33} (1991)
                 2460 [Sov. Phys. - Solid State, {\bf 33} (1991) 1384];
                 {\bf 34} (1992) 97 [{\bf 34} (1992) 50].
\bibitem{Kovar} V.A.Kovarskiy, Fiz. Tv. Tela, {\bf 34} (1992) 3549.
                [Sov. Phys. - Solid State, {\bf 34} (1992) 1900].
\bibitem{Sidor} Yu.V.Sidorov, Europhys. Lett. {\bf 10} (1989) 415.
\bibitem{GrishSid} L.P.Grishchuk and Yu.V.Sidorov, Phys. Rev. D{\bf 42}
                   (1990) 3413.
\bibitem{GrishHaus} L.Grishchuk, H.A.Haus, and K.Bergman, Phys. Rev.
                   D{\bf 46} (1992) 1440.
\bibitem{Albr} A.Albrecht, P.Ferreira, M.Joyce, and T.Prokopec,
              Phys. Rev. D{\bf 50} (1994) 4807.
\bibitem{VourdasWeiner} A.Vourdas and R.M.Weiner, Phys. Rev. D{\bf 38}
                        (1988) 2209.
\bibitem{Ruijgrok} T.W.Ruijgrok, Acta Phys. Pol. B{\bf 23} (1992) 629.
\bibitem{Belorus} S.Ya.Kilin, V.I.Kuvshinov, and S.A.Firago,
                  in Proceedings of the Second International Workshop on
                  Squeezed States and Uncertainty Relations (Moscow,
                  25-29 May, 1992) eds. D.Han, Y.S.Kim, and V.I.Man'ko,
                  NASA Conference Publication No.3219 (1993) p.301.
\bibitem{Dremin} I.M.Dremin, Mod. Phys. Lett. A{\bf 8} (1993) 2747.
\bibitem{DreminPhysLett} I.M.Dremin, Phys. Lett. B{\bf 313} (1993) 209; \\
                   I.M.Dremin and R.C.Hwa, Phys. Rev. D{\bf 49} (1994) 5805.
\bibitem{DreminUFN} I.M.Dremin, Usp. Fiz. Nauk {\bf 164} (1994) 785
                    [Physics --Uspekhi, to appear].
\bibitem{Gian} I.M.Dremin, V.Arena, G.Boca, et al., Phys. Lett. B{\bf 336}
               (1994) 119.
\bibitem{Pavel} V.V.Dodonov, I.M.Dremin, P.G.Polynkin, and V.I.Man'ko,
                Phys. Lett. A{\bf 193} (1994) 209.
\bibitem{Ahiezer} V.B.Berestetskii, E.M.Lifshitz, and L.P.Pitaevskii,
                  ``Quantum Electrodynamics'', Pergamon Press, 1982.
\bibitem{200T} V.V.Dodonov, A.B.Klimov, and V.I.Man'ko,
               in ``Squeezed and Correlated States of Quantum Systems,''
               ed. M.A.Markov, Proceedings of Lebedev Physical Institute,
               vol. 200( Nauka, Moscow, 1991) p.56 [translation: Nova
               Science, Commack, N. Y., 1993, vol.205, p.61].
\bibitem{deGroot} S.R.DeGroot and L.G.Suttorp, Foundations of
                  Electrodynamics (North--Holland, Amsterdam, 1972).
\bibitem{Scully} M.Hillery, R.F.O'Connell, M.O.Scully,
                 and E.P.Wigner, Phys. Rep. {\bf 106} (1984) 123.
\bibitem{Vourdas} A.Vourdas, Phys. Rev. A{\bf 34} 3466 (1986).
\bibitem{[2]}  G.S.Agarwal, J. Mod. Opt. {\bf 34} (1987) 909.
\bibitem{[3]} G.S.Agarwal and G.Adam, Phys. Rev. A{\bf 38} (1988) 750.
\bibitem{[4]} H.Fearn and M.J.Collett, J. Mod. Opt. {\bf 35} (1988) 553.
\bibitem{[5]} G.S.Agarwal and G.Adam, Phys. Rev. A{\bf 39} (1989) 6259.
\bibitem{[6]} M.S.Kim,  F.A.M.~de~Oliveira,  and P.L.Knight,
              Phys. Rev. A{\bf 40} (1989) 2494.
\bibitem{[7]} S.Chaturvedi
              and V.Srinivasan, Phys. Rev. A{\bf 40} (1989) 6095.
\bibitem{[8]} G.Adam, Phys. Lett. A{\bf 171} (1992) 66.
\bibitem{[9]} P.Marian and T.A.Marian, Phys. Rev. A{\bf 47} (1993) 4474.
\bibitem{Car} C.M.Caves, Chang Zhu, G.L.Milburn, and W.Schleich,
                  Phys. Rev. A{\bf 43} (1991) 3854.
\bibitem{sch} G.Schrade, V.Akulin, V.I.Man'ko, and W.Schleich,
                  Phys. Rev. A{\bf 48} (1993) 2398.
\bibitem{[15]} V.V.Dodonov, V.I.Man'ko, and V.V.Semjonov, Nuovo
                    Cimento B{\bf 83} (1984) 145;
                    see also: V.I.Man'ko, in Proceedings of
                    Workshop ``Harmonic Oscillators''
                    (University of Maryland, 25-28 March, 1992) eds. D.Han
                    and Y.S.Kim, NASA Conference Publication.
\bibitem{[14]} Bateman Manuscript Project: Higher Transcendental Functions,
              ed. A.Erd\'elyi (McGraw-Hill, New York, 1953).
\bibitem{167} V.V.Dodonov and V.I.Man`ko,
    in ``Group Theory, Gravitation and Elementary Particle Physics,''
    ed. A.A.Komar, Proceedings of Lebedev Physical Institute, vol.167,
    (Nauka, Moscow, 1986) p.7 [translation: Nova Science, Commack,
    N.Y., 1987, p.7].
\bibitem{[19]} K.B.Wolf, J. Math. Phys. {\bf 15} (1974) 1295; P.Kramer,
               M.Moshinsky, and T.H.Seligman, in ''Group Theory and its
               Applications,'' ed. E.M.Loebl (Academic, New York, 1975)
               p.249.
\bibitem{[20]} E.C.G.Sudarshan,
                      in Proceedings of the Second International Workshop on
                      Squeezed States and Uncertainty Relations (Moscow,
                      25-29 May, 1992) eds. D.Han, Y.S.Kim, and V.I.Man'ko,
                      NASA Conference Publication No.3219 (1993) p.241.

\end{thebibliography}
\end{document}